\documentclass[11pt]{article} 

\usepackage{a4wide}

\RequirePackage[latin1]{inputenc}     
 
\usepackage[english]{babel} 
\usepackage{amsmath} 
\usepackage{amsfonts} 
\usepackage{xspace} 
\usepackage{latexsym} 
\usepackage{url} 
\usepackage{xspace} 
\usepackage[all]{xy} 
\usepackage{graphics,color}              
\RequirePackage[latin1]{inputenc}   
\usepackage{array}
 
\newenvironment{restate-proposition}[2][{}]{\noindent\textbf{Proposition~{#2}}\;\textbf{#1}\  
}{\vskip 1em} 
 
\newenvironment{restate-theorem}[2][{}]{\noindent\textbf{Theorem~{#2}}\;\textbf{#1}\  
}{\vskip 1em} 
 
\newenvironment{restate-corollary}[2][{}]{\noindent\textbf{Corollary~{#2}}\;\textbf{#1}\  
}{\vskip 1em}

\newcommand{\Proofitem}[1]{\medskip \noindent $#1\;$} 
\newcommand{\Proofitemf}[1]{\noindent $#1\;$} 
\newcommand{\Defitem}[1]{\smallskip \noindent $#1\;$} 
 
\newcommand{\Defitemf}[1]{\noindent $#1\;$}

\newcommand{\hbra}{\noindent\hbox to \textwidth{\leaders\hrule height1.8mm depth-1.5mm\hfill}} 
\newcommand{\hket}{\noindent\hbox to \textwidth{\leaders\hrule height0.3mm\hfill}} 
\newcommand{\ratio}{.3}

  \newtheorem{theorem}{Theorem} 
   
  \newtheorem{definition}[theorem]{Definition} 
  \newtheorem{lemma}[theorem]{Lemma} 
   
  \newtheorem{proposition}[theorem]{Proposition}

  \newtheorem{remark}[theorem]{Remark}

 \newcommand{\qed}{\hfill${\Box}$}

\newcommand{\Figbar}{{\center \rule{\hsize}{0.3mm}}}    
 
\newcommand{\cl}[1]{{\cal #1}}          

\newcommand{\Gives}{\vdash}             

\newcommand{\arrow}{\rightarrow}        
\newcommand{\trarrow}{\stackrel{*}{\rightarrow}}        
 
\newcommand{\Nat}{\mathbf{N}}                 
\newcommand{\Alt}{ \mid\!\!\mid  } 
 
\newcommand{\infer}[2]{\begin{array}{c} #1 \\ \hline #2 \end{array}} 
 
\newcommand{\comp}{\circ}               
 
\newcommand{\w}[1]{{\it #1}}    

\newcommand{\xst}[2]{\exists\, #1\;\: #2}

\newcommand{\s}[1]{{\sf #1}}    

\newcommand{\act}[1]{\xrightarrow{#1}} 

\newcommand{\set}[1]{\{#1\}}

 \newcommand{\wact}[1]{\stackrel{#1}{\Rightarrow}} 

\newcommand{\eval}{\Downarrow}

\newcommand{\Z}{{\bf Z}}

\newcommand{\real}{\makebox[5mm]{\,$\|\!-$}}

\newcommand{\imp}{{\sf Imp}}            
\newcommand{\vm}{{\sf Vm}}              
\newcommand{\mips}{{\sf Mips}}          
\newcommand{\C}{{\sf C}}                
\newcommand{\Clight}{{\sf Clight}}        
\newcommand{\Cminor}{{\sf Cminor}}
\newcommand{\RTLAbs}{{\sf RTLAbs}}
\newcommand{\RTL}{{\sf RTL}}
\newcommand{\ERTL}{{\sf ERTL}}
\newcommand{\LTL}{{\sf LTL}}
\newcommand{\LIN}{{\sf LIN}}
\newcommand{\access}[1]{\stackrel{#1}{\leadsto}}
\newcommand{\ocaml}{{\sf ocaml}}
\newcommand{\coq}{{\sf Coq}}
\newcommand{\compcert}{{\sf CompCert}}
\newcommand{\cerco}{{\sf CerCo}}
\newcommand{\cil}{{\sf CIL}}
\newcommand{\scade}{{\sf Scade}}
\newcommand{\absint}{{\sf AbsInt}}
\newcommand{\framac}{{\sf Frama-C}}
\newcommand{\powerpc}{{\sf PowerPc}}
\newcommand{\lustre}{{\sf Lustre}}

\newcommand{\codeex}[1]{\texttt{#1}}   

\begin{document}

\title{Certifying cost annotations in compilers\thanks{This work was supported by the {\em Information and Communication Technologies (ICT) Programme} as Project FP7-ICT-2009-C-243881 $\cerco$.}}

\author{Roberto M. Amadio$^{(1)}$~~~Nicolas~Ayache$^{(2)}$\\
        Yann~R\'egis-Gianas$^{(2)}$ ~~~Ronan~Saillard$^{(2)}$ \\ \\
{\small $^{(1)}$ Universit\'e Paris Diderot (UMR-CNRS 7126)}\\
{\small $^{(2)}$ Universit\'e Paris Diderot (UMR-CNRS 7126) and INRIA (Team ${\pi}r^2$)}}

\maketitle 

\begin{abstract}
We discuss the problem of building a compiler which can {\em lift} in
a provably correct way pieces of information on the execution cost of
the object code to cost annotations on the source code.  To this end,
we need a clear and flexible picture of: (i) the meaning of cost
annotations, (ii) the method to prove them sound and precise, and
(iii) the way such proofs can be composed.  We propose a so-called
{\em labelling} approach to these three questions.  As a first step,
we examine its application to a toy compiler. This formal study
suggests that the labelling approach has good compositionality and
scalability properties.  In order to provide further evidence for this
claim, we report our successful experience in implementing and testing
the labelling approach on top of a prototype compiler written in
$\ocaml$ for (a large fragment of) the {\sc C} language.
\end{abstract}


\section{Introduction}\label{intro-sec}
The formal description and certification of software components is
reaching a certain level of maturity with impressing case studies
ranging from compilers to kernels of operating systems.  A
well-documented example is the proof of functional correctness of a
moderately optimizing compiler from a large subset of the $\C$ language
to a typical assembly language of the kind used in embedded systems 
\cite{Leroy09}.

In the framework of the {\em Certified Complexity} ($\cerco$) project~\cite{Cerco10}, we
aim to refine this line of work by focusing on the issue of the {\em
execution cost} of the compiled code.  Specifically, we aim to build a
formally verified $\C$ compiler that given a source program produces
automatically a functionally equivalent object code plus an annotation
of the source code which is a sound and precise description
of the execution cost of the object code.

We target in particular the kind of $\C$ programs produced for embedded
applications; these programs are eventually compiled to binaries
executable on specific processors.  The current state of the art in
commercial products such as $\scade$~\cite{SCADE,Fornari10} is that the {\em reaction
time} of the program is estimated by means of abstract interpretation
methods (such as those developed by $\absint$~\cite{AbsInt,AbsintScade}) that operate on the
binaries.  These methods rely on a specific knowledge of the
architecture of the processor and may require explicit annotations of
the binaries to determine the number of times a loop is iterated (see,
{\em e.g.},~\cite{W09} for a survey of the state of the art).

In this context, our aim is to produce a functionally correct compiler
which can {\em lift} in a provably correct way the pieces of information on the
execution cost of the binary code to cost annotations on the source $\C$
code.  Eventually, we plan to manipulate the cost annotations 
with automatic tools such as $\framac$~\cite{Frama-C}.
In order to carry on our project, we need a clear and flexible picture
of: (i) the meaning of cost annotations, (ii) the method to prove them
sound and precise, and (iii) the way such proofs can be composed.  Our
purpose here is to propose a methodology addressing these three
questions and to consider its concrete application to a simple toy
compiler and to a moderately optimizing untrusted $\C$ compiler.

\paragraph{Meaning of cost annotations}
The execution cost of the source programs we are interested in depends on
their control structure. Typically, the source programs are composed
of mutually recursive procedures and loops and their execution cost
depends, up to some multiplicative constant, on the number of times
procedure calls and loop iterations are performed.
Producing a {\em cost annotation} of a  source program  amounts to:
\begin{itemize}

\item enrich the program with a  collection of {\em global cost variables} 
to measure resource consumption (time, stack size, heap size,$\ldots$)

\item inject suitable code at some critical points (procedures, loops,$\ldots$) to keep track of the execution cost.  

\end{itemize}

Thus producing a cost-annotation of a source program $P$ amounts to
build an {\em annotated program} $\w{An}(P)$ which behaves as $P$
while self-monitoring its execution cost. In particular, if we do {\em
  not} observe the cost variables then we expect the annotated
program $\w{An}(P)$ to be functionally equivalent to $P$.  Notice that
in the proposed approach an annotated program is a program in the
source language. Therefore the meaning of the cost
  annotations is automatically defined by the semantics of the source
language and tools developed to reason on the source programs 
can be directly applied to the annotated programs too.

\paragraph{Soundness and precision of cost annotations}
Suppose we have a functionally correct compiler $\cl{C}$ that
associates with a program $P$ in the source language a program
$\cl{C}(P)$ in the object language.  Further suppose we have some
obvious way of defining the execution cost of an object code. For
instance, we have a good estimate of the number of cycles needed for
the execution of each instruction of the object code.  Now the
annotation of the source program $\w{An}(P)$ is {\em sound} if its
prediction of the execution cost is an upper bound for the `real'
execution cost. Moreover, we say that the annotation is {\em precise}
with respect to the cost model if the {\em difference} between the
predicted and real execution costs is bounded by a constant which
depends on the program.

\paragraph{Compositionality}\label{comp-intro}
In order to master the complexity of the compilation process (and its
verification), the compilation function $\cl{C}$ must be regarded as 
the result of the composition of a certain number of program transformations
$\cl{C}=\cl{C}_{k} \comp \cdots \comp \cl{C}_{1}$. 
When building a system of
cost annotations on top of an existing compiler  
a certain number of problems arise.
First, the estimated cost of executing a piece of source code
is determined only at the {\em end} of the compilation process.
Thus while we are used to define the compilation 
functions $\cl{C}_i$ in increasing order (from left to right),
the annotation function $\w{An}$ is the result of 
a progressive abstraction from 
the object to the source code (from right to left).
Second, we must be able to foresee in the source language 
the looping and branching  points of the object code. 
Missing a loop may lead to unsound cost annotations
while missing a branching point may lead to rough cost
predictions. This means that we must have a rather good
idea of the way the source code will eventually be compiled
to object code.
Third, the definition of the annotation of the source 
code depends heavily on {\em contextual information}. For instance,
the cost of the compiled code associated with 
a simple expression such as $x+1$ will depend on the
place in the memory hierarchy where the variable $x$ is allocated.
A previous experience described in~\cite{CercoDeliverable} suggests that the process
of pushing `hidden parameters' in the definitions of cost annotations
and of manipulating directly numerical cost is error prone and 
produces complex proofs. For this reason, we advocate next a 
`labelling approach' where costs are handled at
an abstract level and numerical values are produced at the very end of the
construction.

\paragraph{Labelling approach to cost annotations}\label{label-intro}
The `labelling' approach to the problem of building
cost annotations is summarized in the following diagram.













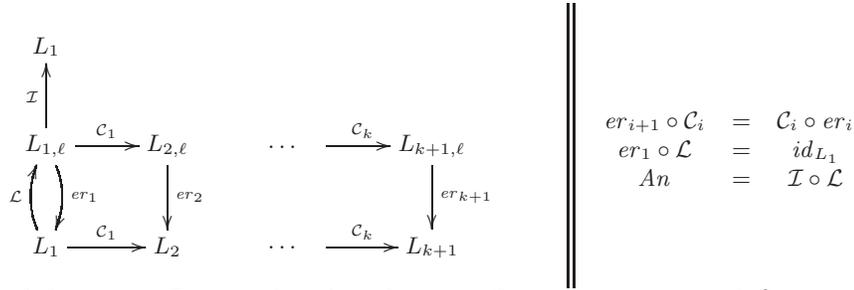
\newcolumntype{M}[1]{>{\raggedright}m{#1}}
\-
{\footnotesize

\begin{tabular}{M{8cm}||M{3cm}}
$$
\xymatrix{
  L_1 \\
%
  L_{1,\ell} 
  \ar[u]^{\cl{I}}
  \ar@/^/[d]^{\w{er}_1} 
  \ar[r]^{\cl{C}_1} 
& L_{2,\ell} 
  \ar[d]^{\w{er}_2}  
& \ldots \hspace{0.3cm}\ar[r]^{\cl{C}_k} 
& L_{k+1, \ell} 
  \ar[d]^{\w{er}_{k+1}}  \\
%
  L_1                                  
  \ar@/^/[u]^{\cl{L}} 
  \ar[r]^{\cl{C}_1}
& L_2   
& \ldots\hspace{0.3cm}
  \ar[r]^{\cl{C}_{k}}
& L_{k+1}
}
$$
&
$$
\begin{array}{ccc}
\w{er}_{i+1} \comp \cl{C}_{i} &= &\cl{C}_{i} \comp \w{er}_{i} \\

\w{er}_{1} \comp \cl{L} &= &\w{id}_{L_{1}} \\ 

\w{An} &= &\cl{I} \comp \cl{L}  
\end{array}
$$
\end{tabular}
}
\-

For each language $L_i$ considered in the compilation process,
we define an extended {\em labelled} language $L_{i,\ell}$ and an
extended operational semantics. The labels are used to mark
certain points of the control. The semantics makes sure 
that whenever we cross a labelled control point a labelled and observable
transition is produced.

For each labelled language there is an obvious function $\w{er}_i$
erasing all labels and producing a program in the corresponding
unlabelled language.
The compilation functions $\cl{C}_i$ are extended from the unlabelled
to the labelled language so that they enjoy commutation 
with the erasure functions. Moreover, we lift the soundness properties of
the compilation functions from the unlabelled to the labelled languages
and transition systems.

A {\em labelling} $\cl{L}$  of the source language $L_1$ is just a function 
such that $\w{er}_{L_{1}} \comp \cl{L}$ is the identity function.
An {\em instrumentation} $\cl{I}$ of the source labelled language  $L_{1,\ell}$
is a function replacing the labels with suitable increments of, say, 
a fresh global \w{cost} variable. Then an {\em annotation} $\w{An}$ of the 
source program can be derived simply as the composition of the labelling
and the instrumentation functions: $\w{An} = \cl{I}\comp \cl{L}$.

Suppose $s$ is some adequate representation
of the state of a program. Let $P$ be a source program and suppose
that its annotation satisfies the following property:
\begin{equation}\label{STEP1}
(\w{An}(P),s[c/\w{cost}])  \eval s'[c+\delta/\w{cost}]
\end{equation}
where $c$ and $\delta$ are some non-negative numbers.
Then the definition of the instrumentation and the fact that
the soundness proofs of the compilation functions have been lifted
to the labelled languages allows to conclude that
\begin{equation}\label{step2}
(\cl{C}(\cl{L}(P)),s[c/\w{cost}])  \eval (s'[c/\w{cost}],\lambda)
\end{equation}
where $\cl{C} = \cl{C}_{k} \comp \cdots \comp \cl{C}_{1}$ and
$\lambda$ is a sequence (or a multi-set) of labels whose `cost'
corresponds to the number $\delta$ produced by the 
annotated program.
Then the commutation properties of erasure and compilation functions
allows to conclude that the {\em erasure} of the compiled
labelled code $\w{er}_{k+1}(\cl{C}(\cl{L}(P)))$ is actually  equal to 
the compiled code $\cl{C}(P)$ we are interested in.
Given this, the following question arises: 
under which conditions 
the sequence $\lambda$, {\em i.e.}, the increment $\delta$,
is a sound and possibly precise description of the
execution cost of the object code? 

To answer this question, we observe that the object code we are interested in is some kind of assembly code 
and its control flow can be easily represented as a control flow
graph. The fact that we have to prove the soundness of the compilation
functions means that we have plenty of information
on the way the control flows in the compiled code, in particular as
far as procedure calls and returns are concerned.
These pieces of information allow to build a rather accurate representation
of the control flow of the compiled code at run time.

The idea is then to perform two simple checks on the control flow
graph. The first check is to verify that all loops go through a labelled node.
If this is the case then we can associate a finite cost with
every label and prove that the cost annotations are sound.
The second check amounts to verify that all paths starting from 
a label have the same cost. If this check is successful then we can
conclude that the cost annotations are precise.

\paragraph{A toy compiler}\label{toy-intro}
As a first case study for the labelling approach to cost
annotations we have sketched, we introduce a
{\em toy compiler} which is summarised by the
following diagram.





\[
\xymatrix{
\imp 
\ar[r]^{\cl{C}}
& \vm 
\ar[r]^{\cl{C'}}
& \mips
}
\]

The three languages considered can be shortly described as follows:
$\imp$ is a very simple imperative language
with pure expressions, branching and looping commands,
$\vm$ is an assembly-like language enriched with a stack, and
$\mips$ is a $\mips$-like assembly language with
registers and main memory.
The first compilation function $\cl{C}$ relies on the stack 
of the $\vm$ language to implement expression evaluation while
the second compilation function $\cl{C'}$ allocates (statically)
the base of the stack in the registers and the rest in main memory.
This is of course a naive strategy but it suffices to expose
some of the problems that arise in defining a compositional approach.

\paragraph{A C compiler}\label{Ccompiler-intro}
As a second, more complex, case study we consider a 
$\C$ compiler we have built in $\ocaml$ whose
structure is summarised by the following diagram:

{\footnotesize
\[
\begin{array}{cccccccccc}
&&\C &\arrow &\Clight &\arrow &\Cminor &\arrow &\RTLAbs &\qquad \mbox{(front end)}\\
                                              &&&&&&&&\downarrow \\
 \mips &\leftarrow &\LIN &\leftarrow  &\LTL &\leftarrow  &\ERTL  &\leftarrow &\RTL &\qquad \mbox{(back-end)}
\end{array}
\]
}

The structure follows rather closely the one of the $\compcert$
compiler~\cite{Leroy09}.  
Notable differences are that some compilation steps are
fusioned, that the front-end goes till $\RTLAbs$ (rather than
$\Cminor$) and that we target the $\mips$ assembly language (rather
than $\powerpc$).  These differences are contingent to the way we
built the compiler.  The compilation from $\C$ to $\Clight$ relies on
the $\cil$ front-end~\cite{CIL02}. The one from $\Clight$ to $\RTL$
has been programmed from scratch and it is partly based on the $\coq$
definitions available in the $\compcert$ compiler.  Finally, the
back-end from $\RTL$ to $\mips$ is based on a compiler developed in
$\ocaml$ for pedagogical purposes~\cite{Pottier}.  The main
optimisations it performs are common subexpression elimination,
liveness analysis and register allocation, and graph compression. We
ran some benchmarks to ensure that our prototype implementation is
realistic. The results are given in
appendix~\ref{C-compiler-benchmarks} and the compiler is available
from the authors.

\paragraph{Organisation}
The rest of the paper is organised as follows.
Section \ref{toy-compiler-sec} describes the 3 languages and the 
2 compilation steps of the toy compiler. 
Section \ref{label-toy-sec} describes the application of
the labelling approach to the toy compiler.
Section \ref{C-label-sec} reports our experience in implementing
and testing the labelling approach on the $\C$ compiler.
Section \ref{conclusion-sec} summarizes our contribution and 
outlines some perspectives for future work.
Appendix \ref{paper-proofs} sketches the proofs that have not
been mechanically checked in $\coq$ and
appendix \ref{C-compiler-sec}  provides some details on the
structure of the $\C$ compiler we have implemented.

\section{A toy compiler}\label{toy-compiler-sec}
We formalise the toy compiler
introduced in section \ref{toy-intro}.

\subsection{\imp: language and semantics}\label{imp-sec}
The syntax of the $\imp$ language is described below.
This is a rather standard imperative language with 
\s{while} loops and \s{if-then-else}. 

{\footnotesize
\[
\begin{array}{lll}

\w{id}&::= x\Alt y\Alt \ldots       &\mbox{(identifiers)} \\
\w{n} &::= 0 \Alt -1 \Alt +1 \Alt \ldots &\mbox{(integers)} \\
v &::= n \Alt \s{true} \Alt \s{false}  &\mbox{(values)} \\
e &::= \w{id} \Alt n \Alt e+e    &\mbox{(numerical expressions)} \\
b &::= e<e  &\mbox{(boolean conditions)} \\
S &::= \s{skip} \Alt \w{id}:=e \Alt S;S \Alt 
  \s{if} \ b \ \s{then} \ S \ \s{else} \ S 
   \Alt \s{while} \ b \ \s{do} \ S  &\mbox{(commands)} \\
P &::= \s{prog} \ S      &\mbox{(programs)}

\end{array}
\]}

Let $s$ be a total function from identifiers to integers representing
the {\bf state}.  If $s$ is a state, $x$ an identifier, and $n$ an
integer then $s[n/x]$ is the `updated' state such that $s[n/x](x)=n$
and $s[n/x](y)=s(y)$ if $x\neq y$.
%
The {\em big-step} operational semantics of $\imp$ 
expressions and boolean conditions is defined as follows:

{\footnotesize
\[
\begin{array}{c}

\infer{}
{(v,s)\eval v}

\qquad

\infer{}
{(x,s) \eval s(x)}

\qquad

\infer{(e,s)\eval v\quad (e',s)\eval v'}
{(e+e',s)\eval (v+_{\Z} v')} 

\qquad
\infer{(e,s)\eval v\quad (e',s)\eval v'}
{(e<e',s) \eval (v<_{\Z} v')} 
\end{array}\]}

A {\em continuation} $K$ is a list of commands which terminates with a special
symbol \s{halt}:
$K::= \s{halt} \Alt S \cdot K$.
Table \ref{small-step-imp} defines a small-step semantics of $\imp$ commands 
whose basic judgement has the shape:
$(S,K,s) \arrow  (S',K',s')$.
We define the semantics of a program $\s{prog} \ S$ as the semantics
of the command $S$ with continuation $\s{halt}$.
We derive a big step semantics from the small step one as follows:
$(S,s) \eval s'$  if $(S,\s{halt},s) \arrow \cdots \arrow (\s{skip},\s{halt},s')$.

\begin{table}
{\footnotesize
\[
\begin{array}{lll}
(x:=e,K,s) &\arrow &(\s{skip},K,s[v/x]) \qquad\mbox{if }(e,s)\eval v \\ \\

(S;S',K,s) &\arrow &(S,S'\cdot K,s) \\ \\

(\s{if} \ b \ \s{then} \ S \ \s{else} \ S',K,s) &\arrow 
&\left\{
\begin{array}{ll}
(S,K,s) &\mbox{if }(b,s)\eval \s{true} \\
(S',K,s) &\mbox{if }(b,s)\eval \s{false}
\end{array}
\right. \\ \\

(\s{while} \ b \ \s{do} \ S ,K,s) &\arrow 
&\left\{
\begin{array}{ll}
(S,(\s{while} \ b \ \s{do} \ S)\cdot K,s) &\mbox{if }(b,s)\eval \s{true} \\
(\s{skip},K,s) &\mbox{if }(b,s)\eval \s{false}
\end{array}
\right. \\ \\

(\s{skip},S\cdot K,s) &\arrow
&(S,K,s)

\end{array}
\]}
\caption{Small-step operational semantics of $\imp$ commands}\label{small-step-imp}
\end{table}

\subsection{\vm: language and semantics}\label{vm-sec}
Following~\cite{Leroy09-ln},
we define a virtual machine $\vm$ and its programming language.
The machine includes the following elements:
(1) a fixed code $C$ (a possibly empty sequence of instructions),
(2) a program counter \w{pc},
(3) a store $s$ (as for the source program),
(4) a stack of integers $\sigma$.



Given a sequence $C$, we denote with $|C|$ its length and 
with  $C[i]$ its $i^{\w{th}}$ element (the leftmost element
being the $0^{\w{th}}$ element).
The operational semantics of the instructions is formalised by rules of the shape
$C \Gives (i,\sigma,s) \arrow (j,\sigma',s')$
and it is fully described in table \ref{semantics-vm}.
Notice that $\imp$ and $\vm$ semantics share the same notion of store.
We write, {\em e.g.}, $n\cdot \sigma$ to stress that the top element 
of the stack exists and is $n$.
We will also write $(C,s)\eval s'$ if 
$C\Gives (0,\epsilon,s) \trarrow (i,\epsilon,s')$ and $C[i]=\s{halt}$.
\begin{table}
{\footnotesize
\[
\begin{array}{l|l}

\mbox{Rule}                                                 & C[i]= \\\hline

C \Gives (i, \sigma, s) \arrow (i+1, n\cdot \sigma,s)  & {\tt cnst}(n) \\ 
C \Gives (i, \sigma, s) \arrow (i+1,s(x)\cdot \sigma,s)  & {\tt var}(x) \\ 
C \Gives (i, n \cdot \sigma, s) \arrow (i+1,\sigma,s[n/x])  & {\tt setvar}(x) \\
C \Gives (i, n \cdot n' \cdot \sigma, s) \arrow (i+1, (n+_{\Z} n')\cdot \sigma,s)  & {\tt add} \\
C \Gives (i, \sigma, s) \arrow (i+k+1, \sigma,s)  & {\tt branch(k)} \\
C \Gives (i, n \cdot n' \cdot \sigma, s) \arrow (i+1, \sigma,s)  & {\tt bge(k)} \mbox{ and }n<_{\Z} n'\\
C \Gives (i, n \cdot n' \cdot \sigma, s) \arrow (i+k+1, \sigma,s)  & {\tt bge(k)} \mbox{ and }n\geq_{\Z} n'\\

\end{array}
\]}
\caption{Operational semantics $\vm$ programs}\label{semantics-vm}
\end{table}

Code coming from the compilation of $\imp$ programs has specific
properties that are used in the following compilation step when values
on the stack are allocated either in registers or in main memory.  In
particular, it turns out that for every instruction of the compiled
code it is possible to predict statically the {\em height of the stack}
whenever the instruction is executed.  We now proceed to define a simple notion
of {\em well-formed} code and show that it enjoys this property.  In
the following section, we will define the compilation function from
$\imp$ to $\vm$ and show that it produces well-formed code.

\begin{definition}
We say that a sequence of instructions $C$ is well formed if there
is a function $h:\set{0,\ldots,|C|} \arrow \Nat$ which satisfies the
conditions listed in table \ref{well-formed-instr} for $0\leq i \leq |C|-1$.
In this case we write $C:h$.
\end{definition}

\begin{table}
{\footnotesize
\[
\begin{array}{l|l}

C[i]=         & \mbox{Conditions for }C:h \\\hline

\s{cnst}(n)
\mbox{ or } \s{var}(x)
&h(i+1) = h(i)+1 \\

\s{add}            
& h(i)\geq 2, \quad h(i+1)=h(i)-1 \\

\s{setvar}(x)      
& h(i)=1, \quad h(i+1)=0 \\

\s{branch}(k)
& 0\leq i+k+1 \leq |C|, \quad h(i)=h(i+1)=h(i+k+1)=0 \\

\s{bge}(k)
& 0\leq i+k+1 \leq |C|, \quad h(i)=2, \quad h(i+1)=h(i+k+1)=0 \\

\s{halt}    
&i=|C|-1, \quad h(i)=h(i+1)=0

\end{array}
\]}
\caption{Conditions for well-formed code}\label{well-formed-instr}
\end{table}

The conditions defining the predicate $C:h$ are strong enough 
to entail that $h$ correctly predicts the stack height 
and to guarantee the uniqueness of $h$ up to the initial condition.

\begin{proposition}\label{unicity-height-prop}
(1) 
If $C:h$, $C\Gives (i,\sigma,s) \trarrow (j,\sigma',s')$, and 
$h(i)=|\sigma|$ then $h(j)=|\sigma'|$.
(2) If $C:h$, $C:h'$ and $h(0)=h'(0)$ then $h=h'$.
\end{proposition}

\subsection{Compilation from $\imp$ to $\vm$}\label{compil-imp-vm-sec}
In table \ref{compil-imp-vm}, 
we define compilation functions $\cl{C}$ from $\imp$ to $\vm$ which
operate on expressions, boolean conditions, statements, and programs.
We write $\w{sz}(e)$, $\w{sz}(b)$, $\w{sz}(S)$ for the number of instructions
the compilation function associates with the expression $e$, the boolean
condition $b$, and the statement $S$, respectively.

\begin{table}
{\footnotesize
\[
\begin{array}{c}

\cl{C}(x) = {\tt var}(x)   \qquad
\cl{C}(n)  ={\tt cnst}(n)   \qquad
\cl{C}(e+e') =\cl{C}(e) \cdot \cl{C}(e') \cdot {\tt add} \\ \\

\cl{C}(e<e',k) = \cl{C}(e') \cdot \cl{C}(e) \cdot {\tt bge(k)} \\ \\ 

\qquad  \cl{C}(x:=e) = \cl{C}(e) \cdot {\tt setvar(x)} 

\qquad \cl{C}(S;S') = \cl{C}(S) \cdot \cl{C}(S')\\ \\

\cl{C}(\s{if} \ b \ \s{then} \ S\  \s{else} \ S') = 
\cl{C}(b,k) \cdot \cl{C}(S) \cdot (\s{branch}(k')) \cdot \cl{C}(S')  \\
\mbox{where: } k= \w{sz}(S)+1, \quad k'=\w{sz}(S')

\\ \\

\cl{C}(\s{while} \ b \ \s{do} \  S )
= \cl{C}(b,k)  \cdot \cl{C}(S)  \cdot \s{branch}(k') \\
\mbox{where: } k=\w{sz}(S)+1, \quad k'=-(\w{sz}(b)+\w{sz}(S)+1)\\\\

\cl{C}(\s{prog} \ S) = 
\cl{C}(S) \cdot \s{halt}

\end{array}
\]}

\caption{Compilation from $\imp$ to $\vm$}\label{compil-imp-vm}
\end{table}

We follow~\cite{Leroy09-ln} for 
the proof of soundness of the compilation function for
expressions and boolean conditions
(see also~\cite{MP67} for a
much older reference).

\begin{proposition}\label{soundness-big-step-prop}
The following properties hold:

\Defitem{(1)}
If $(e,s)\eval v$ then 
$C \cdot \cl{C}(e) \cdot C' 
\Gives (i,\sigma,s) \trarrow (j,v\cdot \sigma,s)$
where $i=|C|$ and $j=|C \cdot \cl{C}(e)|$.

\Defitem{(2)}
If $(b,s)\eval \s{true}$ then 
$C \cdot \cl{C}(b,k) \cdot C' 
\Gives (i,\sigma,s) \trarrow (j+k,\sigma,s)$
where  $i=|C|$ and $j=|C \cdot \cl{C}(b,k)|$.

\Defitem{(3)}
If $(b,s)\eval \s{false}$
then $C \cdot \cl{C}(b,k) \cdot C' 
\Gives (i,\sigma,s) \trarrow (j,\sigma,s)$
where  $i=|C|$ and $j=|C \cdot \cl{C}(b,k)|$.

\end{proposition}

Next we focus on the compilation of statements.
We introduce a ternary relation $R(C,i,K)$ which relates 
a $\vm$ code $C$, a number $i\in\set{0,\ldots,|C|-1}$ and a continuation
$K$. The intuition is that relative to the code $C$, 
the instruction $i$ can be regarded as having continuation $K$. 
(A formal definition is available in appendix~\ref{soundness-small-step}.)
We can then state the correctness of the compilation function as follows.

\begin{proposition}\label{soundness-small-step}
If $(S,K,s) \arrow (S',K',s')$ and $R(C,i,S\cdot K)$ then 
$C\Gives (i,\sigma,s) \trarrow (j,\sigma,s')$ and
$R(C,j,S'\cdot K')$.
\end{proposition}

As announced, we can prove that the result of the compilation
is a well-formed code.

\begin{proposition}\label{well-formed-prop}
For any program $P$  there is a unique $h$ such that $\cl{C}(P):h$.
\end{proposition}

\subsection{$\mips$: language and semantics}\label{mips-sec}
We consider a $\mips$-like machine~\cite{Larus05} 
which includes the following elements:
(1) a fixed code $M$ (a sequence of instructions),
(2) a program counter \w{pc},
(3) a finite set of registers including the registers 
$A$, $B$, and $R_0,\ldots,R_{b-1}$, 
and (4)  an (infinite) main memory which maps 
locations to integers.

We denote with $R,R',\ldots$ registers, with $l,l',\ldots$ locations
and with $m,m',\ldots$ memories which  are total functions from
registers and locations to (unbounded) integers.
%
%
We denote with $M$ a list of instructions.
The operational semantics is formalised  in
table \ref{semantics-mips} by rules of the shape
$M \Gives (i,m) \arrow (j,m')$, 
where $M$ is a list of $\mips$ instructions, $i,j$ are natural numbers and $m,m'$ are memories.
We write $(M,m) \eval m'$ if 
$M\Gives (0,m) \trarrow (j,m')$ and $M[j]=\s{halt}$.

\begin{table}
{\footnotesize
\[
\begin{array}{l|l}

\mbox{Rule}                                                 & M[i]= \\\hline

M \Gives (i, m) \arrow (i+1, m[n/R])  & \s{loadi} \ R,n \\ 

M \Gives (i, m) \arrow (i+1,m[m(l)/R])  & \s{load} \ R,l \\ 

M \Gives (i, m ) \arrow (i+1,m[m(R)/l]))  & \s{store} \ R,l \\

M \Gives (i, m) 
\arrow (i+1, m[m(R')+m(R'')/R])  & \s{add}\ R,R',R'' \\

M \Gives (i, m) \arrow (i+k+1, m)  & \s{branch}\ k \\

M \Gives (i, m) \arrow (i+1, m)  & \s{bge}\ R,R',k \mbox{ and }m(R)<_{\Z} m(R')\\

M \Gives (i, m) \arrow (i+k+1, m)  & \s{bge}\ R,R',k \mbox{ and }m(R)\geq_{\Z} m(R')
\end{array}
\]}
\caption{Operational semantics $\mips$ programs}\label{semantics-mips}
\end{table}

\subsection{Compilation from $\vm$ to $\mips$}\label{compil-vm-mips-sec}
In order to compile $\vm$ programs to $\mips$ programs we make the following
hypotheses:
(1) for every $\vm$ program variable $x$ we reserve an address $l_x$, 
(2) for every natural number $h\geq b$, we reserve an address $l_h$ (the addresses $l_x,l_h,\ldots$ are all distinct), and
(3) we store the first $b$ elements of the stack $\sigma$ in the registers 
$R_0,\ldots,R_{b-1}$ and the remaining (if any) at the addresses
$l_b,l_{b+1},\ldots$. 

We say that the memory $m$ represents the stack $\sigma$ and 
the store $s$, and write $m\real \sigma,s$, 
if the following conditions are satisfied:
(1) $s(x) = m(l_x)$, and (2) if $0\leq i < |\sigma|$ then
$\sigma[i] = m(R_i)$ if $i<b$, and 
$\sigma[i] = m(l_i)$ if $i\geq b$.

\begin{table}
{\footnotesize
\[
\begin{array}{l|l}

C[i]=          &\cl{C'}(i,C)= 
\\\hline

\s{cnst}(n)  
& \left\{ \begin{array}{lr}
                   (\s{loadi} \ R_h,n)                                &\mbox{if } h=h(i)<b \\
                   (\s{loadi} \ A,n ) \cdot  (\s{store} \ A, l_h)     &\mbox{otherwise}
                   \end{array} \right. \\

\s{var}(x)
&  \left\{ \begin{array}{lr}
                   (\s{load} \ R_h,l_x)                           &\mbox{if } h=h(i)<b \\
                   (\s{load} \ A,l_x ) \cdot  (\s{store} \ A, l_h)  &\mbox{otherwise}
                   \end{array} \right. \\

\s{add}
& \left\{ \begin{array}{lr}
                     (\s{add} \ R_{h-2} , R_{h-2} , R_{h-1})   &\mbox{if }h=h(i)<(b-1)   \\
                     (\s{load} \ A,l_{h-1}) \cdot (\s{add} \ R_{h-2}, R_{h-2}, A)         &\mbox{if }h=h(i)=(b-1)  \\

                           (\s{load} \ A, l_{h-1}) \cdot (\s{load} \ B,l_{h-2})  &\mbox{if }h=h(i)>(b-1) \\ 
                           (\s{add} \ A,B,A) \cdot (\s{store} \ A, l_{h-2})         
                        \end{array}\right. \\ 

\s{setvar}(x)            
& \left\{ \begin{array}{lr}
                         (\s{store} \ R_{h-1} \ l_x) &\mbox{if }h=h(i)<b\\
                         (\s{load} \ A,l_{h-1})\cdot (\s{store} \ A,l_{x}) &\mbox{if }h=h(i)\geq b
                         \end{array}\right. \\

\s{branch}(k)
& (\s{branch} \ k')  \quad \mbox{if }k'=p(i+k+1,C)-p(i+1,C)\\

\s{bge}(k)
&\left\{ \begin{array}{lr}
                     (\s{bge} \ R_{h-2} , R_{h-1} , k')                                &\mbox{if }h=h(i)<(b-1)   \\

                     (\s{load} \ A,l_{h-1}) \cdot (\s{bge} \ R_{h-2}, A, k')         &\mbox{if }h=h(i)=(b-1)  \\

                           (\s{load} \ A, l_{h-2}) \cdot (\s{load} \ B,l_{h-1}) \cdot 
                           (\s{bge} \ A,B,k')         &\mbox{if }h=h(i)>(b-1), \  k'= \\

&p(i+k+1,C)-p(i+1,C)
          \end{array}\right. \\

\s{halt}
&\s{halt}

\end{array}
\]}
\caption{Compilation from $\vm$ to $\mips$}\label{compil-vm-mips}
\end{table}

The compilation function $\cl{C'}$ from 
$\vm$ to $\mips$ is described in table \ref{compil-vm-mips}.
It operates on a well-formed $\vm$ code $C$ whose last instruction
is $\s{halt}$. Hence, by proposition \ref{well-formed-prop}(3), 
there is a unique $h$ such that $C:h$.
We denote with $\cl{C'}(C)$ the concatenation $\cl{C'}(0,C)\cdots \cl{C'}(|C|-1,C)$.
Given a well formed $\vm$ code $C$ with $i<|C|$ we denote with 
$p(i,C)$ the position of the first instruction in $\cl{C'}(C)$ which
corresponds to the compilation of the instruction with position $i$ in $C$.
This is defined as\footnote{There is an obvious circularity in this definition that can be easily eliminated by defining first the function 
$d$ following the case
analysis in table \ref{compil-vm-mips}, then the function $p$, and finally the function $\cl{C'}$ as in table \ref{compil-vm-mips}.}
$p(i,C) = \Sigma_{0\leq j <i} d(i,C)$,
where the function $d(i,C)$ is defined as 
$d(i,C) = |\cl{C'}(i,C)|$.
Hence $d(i,C)$ is the number of $\mips$ instructions associated
with the $i^{\w{th}}$ instruction of the (well-formed) $C$ code.
The functional correctness of the compilation function can then be stated 
as follows.

\begin{proposition}\label{simulation-vm-mips}
Let $C : h$ be a well formed code.
If  $C \Gives (i,\sigma,s) \arrow (j,\sigma',s')$ with $h(i) = |\sigma|$ and  $m \real \sigma,s$
then 
$\cl{C'}(C) \Gives (p(i,C),m) \trarrow (p(j,C),m')$ and $m'\real \sigma',s'$.
\end{proposition}

\section{Labelling approach for the toy compiler}\label{label-toy-sec}
We apply the labelling approach introduced in section 
\ref{label-intro} to the toy compiler which results in the following
diagram.

\-
{\footnotesize

\newcolumntype{M}[1]{>{\raggedright}m{#1}}
\begin{tabular}{M{7cm}||M{3cm}}
$$
\xymatrix{
  \imp \\
%
  \imp_\ell
  \ar[u]^{\cl{I}}
  \ar@/^/[d]^{\w{er}_\imp} 
  \ar[r]^{\cl{C}} 
& \vm_\ell
  \ar[d]^{\w{er}_\vm}  
  \ar[r]^{\cl{C}'} 
& \mips_\ell
  \ar[d]^{\w{er}_{\mips}}  \\
%
  \imp                
  \ar@/^/[u]^{\cl{L}} 
  \ar[r]^{\cl{C}}
& \vm   
  \ar[r]^{\cl{C}'}
& \mips
}
$$
&
$$
\begin{array}{ccc}
\w{er}_\vm \comp \cl{C} &= &\cl{C} \comp \w{er}_{\imp} \\

\w{er}_\mips \comp \cl{C'} &= &\cl{C'} \comp \w{er}_{\vm} \\

\w{er}_\imp \comp \cl{L} & = & id_{\imp} \\

\w{An}_\imp &= &\cl{I} \comp \cl{L}  
\end{array}
$$
\end{tabular}}















\subsection{Labelled $\imp$}
We extend the syntax so that statements
can be labelled: $S::=  \ldots \Alt \ell:S$.
%
%
%
%
For instance,
$\ell:(\s{while} \ (n<x) \ \s{do} \ \ell:S )$ is a labelled command.
The small step semantics of statements 
defined in table \ref{small-step-imp}
is extended as follows.

{\footnotesize
\[
\begin{array}{lll}
(\ell:S,K,s) &\act{\ell} &(S,K,s) 



\end{array}
\]}

We denote with $\lambda,\lambda',\ldots$
finite sequences of labels. In particular, we denote with $\epsilon$ the empty sequence
and identify an unlabelled transition with a transition labelled with $\epsilon$.
Then the small step reduction relation we have defined
on statements becomes a {\em labelled transition system}.
There is an obvious {\em erasure} function $\w{er}_{\imp}$ 
from the labelled language to the unlabelled one which is the identity on expressions
and boolean conditions, 
and traverses commands removing all labels.
We derive a {\em labelled} big-step semantics as follows:
$(S,s)\eval (s',\lambda)$ if $(S,\s{halt},s) \act{\lambda_1} \cdots \act{\lambda_n}  
                                             (\s{skip},\s{halt},s')$  and $\lambda =\lambda_1 \cdots \lambda_n$.


\subsection{Labelled $\vm$}
We introduce a new instruction
$\s{nop}(\ell)$
whose semantics is defined as follows:

{\footnotesize
\[
C \Gives (i,\sigma,s) \act{\ell} (i+1,\sigma,s) \qquad \mbox{if }C[i]=\s{nop}(\ell)~.
\]}

The erasure function $\w{er}_{\vm}$ amounts to remove from a
$\vm$ code $C$ all the $\s{nop}(\ell)$ instructions and recompute jumps
accordingly. Specifically, let $n(C,i,j)$ be the number of \s{nop} instructions
in the interval $[i,j]$. Then, assuming $C[i]=\s{branch}(k)$ we replace the offset
$k$ with an offset $k'$ determined as follows:

{\footnotesize
\[
k' = 
\left\{
\begin{array}{ll}

k-n(C,i,i+k) &\mbox{if }k\geq 0 \\

k+n(C,i+1+k,i)   &\mbox{if }k<0

\end{array}
\right.
\]}
The compilation function $\cl{C}$ is extended to $\imp_\ell$ by defining:

{\footnotesize
\[
\begin{array}{llll}   
\cl{C}(\ell:b,k) &=(\s{nop}(\ell))\cdot \cl{C}(b,k) \qquad
&\cl{C}(\ell:S)   &=   (\s{nop}(\ell))\cdot \cl{C}(S)~.

\end{array}
\]}

\begin{proposition}\label{labelled-sim-imp-vm}
For all commands $S$ in $\imp_\ell$ we have that:

\Defitem{(1)} $\w{er}_{\vm}(\cl{C}(S)) = \cl{C}(\w{er}_{\imp}(S))$.

\Defitem{(2)} 
If $(S,s)\eval (s',\lambda)$ then $(\cl{C}(S),s)\eval(s',\lambda)$.
\end{proposition}

\begin{remark}
\label{rem:multi-set}
In the current formulation, 
a sequence of transitions $\lambda$ in the source code 
must be simulated by the same sequence of transitions
in the object code. However, in the actual computation
of the costs, the order of the labels occurring in the sequence is
immaterial. Therefore one may consider a more relaxed notion of simulation
where $\lambda$ is a multi-set of labels.
\end{remark}

\subsection{Labelled $\mips$}
The labelled extension of $\mips$ is similar to the one of $\vm$.
We add an instruction $\s{nop} \ \ell$ whose semantics is defined
as follows:

{\footnotesize
\[
\begin{array}{ll}
M\Gives (i,m) \act{\ell} (i+1,m)    &\mbox{if }M[i]=(\s{nop} \ \ell)~.
\end{array}
\]}

The {\em erasure function} $\w{er}_{\mips}$ is also similar to the one 
of $\vm$ as it amounts to remove from a $\mips$ code all the $(\s{nop} \ \ell)$ 
instructions and recompute jumps accordingly. 
The compilation function $\cl{C'}$ is extended to $\vm_\ell$ by 
simply translating $\s{nop}(\ell)$ as $(\s{nop} \ \ell)$:

{\footnotesize
\[
\begin{array}{ll}   

\cl{C'}(i,C) = (\s{nop} \ \ell)
&\mbox{if }C[i]=\s{nop}(\ell)

\end{array}
\]}
The evaluation predicate for labelled $\mips$ is defined as
$(M,m)\eval (m',\lambda)$
if 
$M\Gives (0,m) \act{\lambda_1} \cdots \act{\lambda_n}  (j,m')$, 
$\lambda =\lambda_1 \cdots \lambda_n$  and $M[j]=\s{halt}$.
%
%
The following proposition relates $\vm_\ell$ code and its compilation
and it is similar to proposition \ref{labelled-sim-imp-vm}.

\begin{proposition}\label{sim-vm-mips-prop}
Let $C$ be a $\vm_\ell$ code. Then:

\Defitem{(1)} $\w{er}_{\mips}(\cl{C'}(C)) = \cl{C'}(\w{er}_{\vm}(C))$.

\Defitem{(2)} 
If $(C,s) \eval (s',\lambda)$ and $m\real \epsilon,s$ then
($\cl{C'}(C),m) \eval (m',\lambda)$ and $m'\real \epsilon,s'$.
\end{proposition}

\subsection{Labellings and instrumentations}
Assuming a function $\kappa$ which associates an integer number with
labels and a distinct variable \w{cost} which does not occur in the program $P$ under consideration,
we abbreviate with $\w{inc}(\ell)$ the assignment $\w{cost}:=\w{cost}+\kappa(\ell)$.
Then we define the instrumentation $\cl{I}$ (relative to $\kappa$ and $\w{cost})$ 
as follows:
\[
\cl{I}(\ell:S) = \w{inc}(\ell) ; \cl{I}(S) ~.
\]



The function $\cl{I}$ just distributes over the other operators of the language.
We extend the function $\kappa$ on labels to sequences of labels by defining
$\kappa(\ell_1,\ldots,\ell_n) = \kappa(\ell_1)+\cdots +\kappa(\ell_n)$.
The instrumented $\imp$ program relates to the labelled one has follows.

\begin{proposition}\label{lab-instr-erasure-imp}
Let $S$ be an $\imp_\ell$ command. If 
$(\cl{I}(S),s[c/\w{cost}]) \eval s'[c+\delta/\w{cost}]$
then 
$\xst{\lambda}{\kappa(\lambda)=\delta \mbox{ and }(S,s[c/\w{cost}])\eval(s'[c/\w{cost}],\lambda)}$.
\end{proposition}

\begin{definition}\label{labelling-def}
A {\em labelling} is a function $\cl{L}$ from an unlabelled language to 
the corresponding labelled one such that
$\w{er}_{\imp} \comp \cl{L}$ is the identity function on the 
$\imp$ language.
\end{definition}

\begin{proposition}\label{global-commutation-prop}
For any labelling function $\cl{L}$, and $\imp$ program $P$, the following holds:
\begin{equation}
\w{er}_{\mips}(\cl{C'}(\cl{C}(\cl{L}(P))) = \cl{C'}(\cl{C}(P))~.
\end{equation}
\end{proposition}

\begin{proposition}\label{instrument-to-label-prop}
Given a function $\kappa$ for the labels and a labelling function $\cl{L}$,
for all programs $P$ of the source language if 
$(\cl{I}(\cl{L}(P)),s[c/\w{cost}]) \eval s'[c+\delta/\w{cost}]$ 
and $m\real \epsilon,s[c/\w{cost}]$ then 
$(\cl{C'}(\cl{C}(\cl{L}(P))),m) \eval (m',\lambda)$, 
$m'\real \epsilon,s'[c/\w{cost}]$ and 
$\kappa(\lambda)=\delta$.
\end{proposition}

\subsection{Sound and precise labellings}\label{sound-label-sec}
With any $\mips_\ell$ code $M$ we can associate a directed and rooted 
(control flow) graph whose nodes are the instruction positions $\set{0,\ldots,|M|-1}$,
whose root is the node $0$,  and
whose directed edges correspond to the possible transitions between instructions.  
We say that a node is labelled if it corresponds to an instruction $\s{nop}
\ \ell$.  

\begin{definition}
A {\em simple path} in a $\mips_\ell$ code $M$ is a directed finite path in the graph associated with 
$M$ where the first node is labelled, the last node is the predecessor of either a labelled node or a leaf,
and all the other nodes are unlabelled.
\end{definition}

\begin{definition}
A $\mips_\ell$ code $M$ is {\em soundly labelled} if 
in the associated graph 
the root node $0$ is labelled 
and there are no loops that do not go through a labelled node.
\end{definition}

In a soundly labelled graph there are finitely many simple paths.
Thus, given a soundly labelled $\mips$ code $M$, we can associate 
with every label $\ell$ a number $\kappa(\ell)$ which is the
maximum (estimated) cost  of executing a simple 
path whose first node is labelled with $\ell$. 
We stress that in the following we assume that 
the cost of a simple path is proportional to the number of $\mips$
instructions that are crossed in the path.

\begin{proposition}\label{sound-label-prop}
If $M$ is soundly labelled and $(M,m) \eval (m',\lambda)$ then the cost
of the computation is bounded by $\kappa(\lambda)$.
\end{proposition}

Thus for a soundly labelled $\mips$ code the sequence of labels associated
with a computation is a significant information on the execution cost.

\begin{definition}
We say that a soundly labelled code is {\em precise}
if for every label $\ell$ in the code, the simple 
paths starting from a node labelled with $\ell$ have
the same cost.
\end{definition}

In particular, a code is precise if we can associate at most one simple
path with every label.

\begin{proposition}\label{precise-label-prop}
If $M$ is precisely labelled and $(M,m) \eval (m',\lambda)$ then the cost
of the computation is $\kappa(\lambda)$.
\end{proposition}

The next point we have to check is that 
there are labelling functions (of the source code)
such that the compilation function does produce sound and possibly
precise labelled $\mips$ code.  To discuss this point, we introduce in table
\ref{labelling} two labelling functions $\cl{L}_s$ and $\cl{L}_p$ for the
$\imp$ language. The first labelling relies on just one label while
the second one relies on a function ``\w{new}'' which is meant to return fresh
labels and on an auxiliary function $\cl{L'}_p$ which returns a labelled command
and a binary directive $d\in \set{0,1}$. If $d=1$
then the command that follows (if any) must be labelled.

\begin{table}
{\footnotesize
\[
\begin{array}{ll}
\cl{L}_s(\s{prog} \ S) &=  \s{prog}\  \ell: \cl{L}_s(S) \\

\cl{L}_s(\s{skip})     &=\s{skip} \\

\cl{L}_s(x:=e)         &=x:=e \\

\cl{L}_s(S;S')         &= \cl{L}_s(S);\cl{L}_s(S') \\

\cl{L}_s(\s{if} \ b \ \s{then} \ S_1 \ \s{else} \ S_2) &= 

\s{if} \ b \ \s{then} \ \cl{L}_s(S_1) \ \s{else} \ \cl{L}_s(S_2) \\

\cl{L}_s(\s{while} \ b \ \s{do} \ S) &= \s{while} \ b \ \s{do} \ \ell:\cl{L}_s(S) \\  \\

\cl{L}_p(\s{prog} \ S) &= \s{prog}\   \cl{L}_p(S) \\

\cl{L}_p(S) &= \w{let} \ \ell=\w{new}, \ (S',d) =  \cl{L'}_p(S)  \ \w{in} \ \ell: S' \\

\cl{L'}_p(S)   &=(S,0)\quad \mbox{if }S=\s{skip}\mbox{ or }S=(x:=e) \\

\cl{L'}_p(\s{if} \ b \ \s{then} \ S_1 \ \s{else} \ S_2) 
&= (\s{if} \ b \ \s{then} \ \cl{L}_p(S_1) \ \s{else} \ \cl{L}_p(S_2), 1) \\ 

\cl{L'}_p(\s{while} \ b \ \s{do} \ S) 
&= (\s{while} \ b \ \s{do} \ \cl{L}_p(S), 1) \\

\cl{L'}_p(S_1;S_2)         &= \w{let} \ (S'_1,d_1)= \cl{L'}_p(S_1),  \ (S'_2,d_2)= \cl{L'}_p(S_2) \ \w{in} \ \\ 
                           &\qquad \w{case} \ d_1\\
                           &\qquad 0: (S'_1;S'_2,d_2) \\
                           &\qquad 1: \w{let} \ \ell= \w{new} \ \w{in} \ (S'_1;\ell:S'_2,d_2) 

\end{array}
\]}
\caption{Two labellings for the $\imp$ language} \label{labelling}
\end{table}

\begin{proposition}\label{lab-sound}
For all $\imp$ programs $P$: 

\Defitem{(1)} 
$\cl{C'}(\cl{C}(\cl{L}_s(P))$ 
is a soundly labelled $\mips$ code.

\Defitem{(2)}
$\cl{C'}(\cl{C}(\cl{L}_p(P))$ is a 
soundly and precisely  labelled $\mips$ code.
\end{proposition}

For an example of command which is not soundly labelled, consider
$\ell: \s{while} \ 0<x \ \s{do} \ x:=x+1$, which when compiled, produces
a loop that does not go through any label.  On the other hand, for an
example of a program which is not precisely labelled consider
$\ell:(\s{if} \ 0<x \ \s{then} \ x:=x+1 \ \s{else} \ \s{skip})$.  In the compiled code, we
find two simple paths associated with the label $\ell$ whose cost will
be quite different in general.

Once a sound and possibly precise labelling $\cl{L}$ has been
designed, we can determine the cost of each label and define an instrumentation
$\cl{I}$ whose composition with $\cl{L}$ will produce the desired
cost annotation.

\begin{definition}
Given a labelling function $\cl{L}$ for the source language $\imp$ 
and a program $P$  in the $\imp$ language,  we define an
annotation for the source program as follows:
\[
\w{An}_{\imp}(P) = \cl{I}(\cl{L}(P))~.
\]
\end{definition}

\begin{proposition}\label{ann-correct}
If $P$ is a program and $\cl{C'}(\cl{C}(\cl{L}(P)))$ is a sound (sound and precise)
labelling then $(\w{An}_{\imp}(P),s[c/\w{cost}]) \eval s'[c+\delta/\w{cost}]$
and $m\real \epsilon,s[c/\w{cost}]$ entails that 
$(\cl{C'}(\cl{C}(P)),m) \eval m'$, $m'\real \epsilon,s'[c/\w{cost}]$ 
and the cost of the execution is bound (is exactly) $\delta$.
\end{proposition}

To summarise, producing sound and precise labellings is mainly a
matter of designing the labelled source language so that 
the labelling is sufficiently {\em fine grained}. 
For instance, in the toy compiler, it enough to label commands while it is not necessary to label
boolean conditions and expressions.

Besides soundness and precision, a third criteria to evaluate 
labellings is that they do not introduce too many unnecessary labels. 
We call this property {\em economy}.
There are two reasons for this requirement. On one hand we would like
to minimise the number of labels so that the source program 
is not cluttered by too many cost annotations and 
on the other hand we would like to maximise the length of the simple
paths because in a modern processor the longer the sequence of
instructions we consider the more accurate is the estimation of
their execution cost (on a long sequence certain costs are 
amortized).
In practice, it seems that one can produce first a sound and 
possibly precise labelling and
then apply heuristics to eliminate unnecessary labels.


\section{Labelling approach for the $\C$ compiler}\label{C-label-sec}

This section informally describes the labelled extensions of the languages in 
the compilation chain (see appendix \ref{C-compiler-sec} for details), 
the way the labels are propagated by the compilation
functions, the labelling of the source code, the hypotheses on the control flow
of the labelled $\mips$ code and the verification that we perform on it, the way
we build the instrumentation, and finally the way the labelling approach has
been tested. 
The process of annotating a $\Clight$ program using the labelling approach is
summarized in table \ref{annot-clight-summary} and is detailed in the following sections.

\begin{table}
{\footnotesize

1. Label the input $\Clight$ program.\\

2. Compile the labelled $\Clight$ program in the labelled world. This
  produces a labelled $\mips$ code. \\

3. For each label of the labelled $\mips$ code, compute the cost of the
  instructions under its scope and generate a \emph{label-cost mapping}. An
  unlabelled $\mips$ code --- the result of the compilation --- is obtained by
  removing the labels from the labelled $\mips$ code. \\

4. Add a fresh \emph{cost variable} to the labelled $\Clight$ program and
  replace the labels by an increment of this cost variable according to the
  label-cost mapping. The result is an \emph{annotated} $\Clight$ program with no
  label.\\
}
\caption{Building the annotation of a $\Clight$ program in the labelling approach}\label{annot-clight-summary}
\end{table}

\subsection{Labelled languages} 
Both the $\Clight$ and $\Cminor$ languages are extended in the same way
by labelling both statements and expressions (by comparison, in the
toy language $\imp$ we just labelled statements).
The labelling of expressions aims to capture precisely their execution cost.
Indeed,  $\Clight$ and $\Cminor$ include expressions such 
as $a_1?a_2;a_3$ whose evaluation cost depends on the boolean value $a_1$.
As both languages are extended in the same way, the extended
compilation does nothing more than sending $\Clight$ labelled
statements and expressions to those of $\Cminor$.

The labelled versions of $\RTLAbs$ and the languages in the back-end
simply consist in adding a new instruction whose semantics is to emit a label
without modifying the state. For the CFG based languages
($\RTLAbs$ to $\LTL$), this new instruction is $\s{emit}\ \w{label} \rightarrow
\w{node}$. For $\LIN$ and $\mips$, it is $\s{emit}\ \w{label}$. The translation
of these label instructions is immediate.
In $\mips$, we also 
rely on a reserved label $\s{begin\_function}$  to pinpoint the
beginning of a function code (cf. section \ref{fun-call-sec}).


\subsection{Labelling of the source language}
As for the toy compiler (cf. end of section \ref{label-toy-sec}), the goals of a labelling are
soundness, precision, and possibly economy.
We explain our labelling by considering the constructions of $\Clight$ and their compilation to
$\mips$.

\paragraph{Sequential instructions}
A sequence of $\Clight$ instructions that compile to sequential $\mips$ code,
such as a sequence of assignments, can be handled by a single 
label which covers the unique execution path.

\paragraph{Ternary expressions and conditionals}
Most $\Clight$ expressions compile to sequential $\mips$ code. 
{\em Ternary expressions}, that introduce a branching in the control
flow, are one exception. In this case, we achieve precision by associating a label with each branch.
This is similar to the treatment of the conditional we have already discussed
in section \ref{label-toy-sec}. As for the $\Clight$ operations \texttt{\&\&} and \texttt{||}
which have a lazy semantics, they are transformed to ternary expressions
\emph{before} computing the labelling.

\paragraph{Loops}
Loops in $\Clight$ are guarded by a condition. Following the arguments for the
previous cases, we add two labels when encountering a loop construct: one label
to start the loop's body, and one label when exiting the loop. 
This is similar to the treatment of \s{while} loops discussed in section \ref{label-toy-sec} and it
is enough to guarantee that the loop in the compiled code goes through a label.

\paragraph{Program Labels and Gotos}
In $\Clight$, program labels and gotos are intraprocedural. Their only effect on
the control flow of the resulting assembly code is to potentially introduce an
unguarded loop. This loop must contain at least one cost label in order to
satisfy the soundness condition, which we ensure by adding a cost label right
after a program label.

\begin{center}
  {\footnotesize
    \begin{tabular}{lllll}
      $\Clight$ & $\xrightarrow{Labelling}$ & Labelled $\Clight$ &
      $\xrightarrow{Compilation}$ & Labelled $\mips$\\\\
      \codeex{lbl:} & & \codeex{lbl:} & & \codeex{lbl:}\\
      \codeex{i++;} & & \textbf{\_cost:} & &
      \codeex{emit} \textbf{\_cost}\\
      \codeex{...} & & \codeex{i++;} & &
      \codeex{li\ \ \ \$v0, 1}\\
      \codeex{goto lbl;} & & \codeex{...} & &
      \codeex{add\ \ \$a0, \$a0, \$v0}\\
      & & \codeex{goto lbl;} & & \codeex{...}\\
      & & & & \codeex{j\ \ \ \ lbl}
    \end{tabular}}
\end{center}

\paragraph{Function calls}\label{fun-call-sec}
Function calls in $\mips$ are performed by indirect jumps, the address
of the callee being in a register. In the general case, this address
cannot be inferred statically. Even though the destination point of a
function call is unknown, when the considered $\mips$ code has been
produced by our compiler, we know for a fact that this function ends
with a return statement that transfers the control back to the
instruction following the function call in the caller. As a result, we
treat function calls according to the following global invariants of
the compilation: (1) the instructions of a function are covered by the
labels inside this function, (2) we assume a function call always
returns and runs the instruction following the call.  Invariant (1)
entails in particular that each function must contain at least one
label. To ensure this, we simply add a starting label in every
function definition. The example below illustrates this point:

\begin{center}
  {\footnotesize
    \begin{tabular}{lllll}
      $\Clight$ & $\xrightarrow{Labelling}$ & Labelled $\Clight$ &
      $\xrightarrow{Compilation}$ & Labelled $\mips$\\\\
      \codeex{void f () \{} & & \codeex{void f () \{} & & \codeex{f\_start:}\\
      \codeex{\ \ f}\emph{'s body} & & \textbf{\ \ \_cost:} & &
      \emph{Frame Creation}\\
      \codeex{\}} & & \codeex{\ \ f}\emph{'s body} & & \emph{Initializations}\\
      & & \codeex{\}} & & \codeex{emit} \textbf{\_cost} \\
      & & & & \codeex{f}\emph{'s body}\\
      & & & & \emph{Frame Deletion}\\
      & & & & \codeex{return}
    \end{tabular}}
\end{center}

We notice that some instructions in $\mips$ will be inserted \emph{before}
the first label is emitted. These instructions relate to the
frame creation and/or variable initializations, and are composed 
of sequential instructions (no branching). To deal with
this issue, we take the convention that the instructions
that precede the first label in a function code are actually
under the scope of the first label.
Invariant (2) is of course an over-approximation of the program
behaviour as a function might fail to return because of an infinite loop. 
In this case, the proposed labelling remains correct: it just assumes
that the instructions following the function call will be executed, and takes
their cost into consideration. The final computed cost is still an
over-approximation of the actual cost. 

\subsection{Verifications on the object code}

The labelling previously described has been designed so that the
compiled $\mips$ code satisfies the soundness and precision
conditions. However, we do not need to prove this, instead we have to
devise an algorithm that checks the conditions on the compiled code.
The algorithm assumes a correct management of function calls in the
compiled code. In particular, when we call a function we always jump
to the first instruction of the corresponding code segment and when we
return we always jump to an an instruction that follows a call.  We
stress that this is a reasonable hypothesis that is essentially
subsumed by the proof that the object code {\em simulates} the source
code.

In our current implementation, we check the soundness and the
precision conditions while building at the same time the label-cost
mapping. To this end, the algorithm takes the following main steps.

{\footnotesize
\begin{itemize}
\item First, for each function a control flow graph is built.

\item For each graph, we check whether there is a unique
label that is reachable from the root by a unique path.
This unique path corresponds to the instructions generated by 
the calling conventions as discussed in section \ref{fun-call-sec}.
We shift the occurrence of the label to the root of the graph.

\item 
By a  strongly connected components algorithm,
we check whether every loop in the graphs goes through at least one label.

\item We perform a (depth-first) search of the graph. Whenever we reach a
  labelled node, we perform a second (depth-first) search that stops at labelled
  nodes and computes an upper bound on the cost of the occurrence of the label. Of
  course, when crossing a branching instruction, we take the maximum cost of the
  branches.  When the second search stops we update the current cost of the
  label-cost mapping (by taking a maximum) and we continue the first search.

\item Warning messages are emitted whenever the maximum is taken between two
different values as in this case  the precision condition may be violated.

\end{itemize}}

\subsection{Building the cost annotation}

Once the label-cost mapping is computed, instrumenting the labelled source code is an easy task. A fresh global variable which we call
\emph{cost variable} is added to the source program with the  purpose of holding the cost value and it is initialised at the very beginning of the \codeex{main} program.
Then, every label is replaced by an increment of the 
cost variable according to the label-cost
mapping. 
Following this replacement, the cost labels disappear and the result is
a $\Clight$ program with annotations in the form of assignments.

There is one final problem: labels inside expressions. 
As we already mentioned,  $\Clight$ does not allow 
writing side-effect instructions --- such as  cost
increments --- inside expressions. To cope with this restriction, we produce
first an instrumented $\C$ program --- with side-effects in expressions 
--- that we translate back to $\Clight$ using $\cil$. 
This process is summarized below.

\begin{center}
  {\footnotesize
    \begin{tabular}{lllll}
      $\left . \begin{array}{l}
      \text{Labelled $\Clight$}\\
      \text{label-cost mapping}
    \end{array} \right \}$ & $\xrightarrow{\text{Instrumentation}}$ &
      Instrumented $\C$ & $\xrightarrow{\cil}$ & Instrumented $\Clight$
    \end{tabular}}
\end{center}

\subsection{Testing}
It is desirable to test the coherence of the labelling from $\Clight$ to $\mips$. 
To this end, each labelled language comes with an interpreter
that produces the trace of the labels encountered during the computation.
Then, one naive approach is to test 
the equality of the traces 
produced by the program at the different stages of the compilation.
Our current implementation passes this kind of tests.
For some optimisations that may re-order computations, the weaker
condition mentioned in remark~\ref{rem:multi-set} could be considered.



\section{Conclusion and future work}\label{conclusion-sec}
We have discussed the problem of building a compiler which can {\em
  lift} in a provably correct way pieces of information on the
execution cost of the object code to cost annotations on the source
code.  To this end, we have introduced the so called 
{\em labelling} approach and discussed its formal application to a
toy compiler. Based on this experience, we have argued that the 
approach has good scalability properties, and  to substantiate this
claim, we have reported on our successful experience in implementing
and testing the labelling approach on top of a prototype compiler
written in $\ocaml$ for a large fragment of the $\C$ language which
can be shortly described as $\Clight$ without floating point.

We discuss next a few directions for future work.
First, we are currently testing the current compiler on 
the kind of $\C$ code produced for embedded applications
by a $\lustre$ compiler.
Starting from the annotated $\C$ code, we are relying on 
the $\framac$ tool to produce automatically meaningful information on, say,
the reaction time of a given synchronous program.
Second, we are porting the current compiler to other assembly
languages. In particular, we are interested in targeting
one of the assembly languages covered by the $\absint$ tool so as 
to obtain more realistic estimations of
the execution cost of sequences of instructions.
Third,  we plan to formalise and validate in the {\em Calculus of Inductive
Constructions} the prototype implementation of the labelling approach
for the $\C$ compiler described in section \ref{C-compiler-sec}.
This requires a major implementation effort which will be carried
on in collaboration with our partners of  the $\cerco$ project~\cite{Cerco10}.
%

\newpage
\appendix

\section{Proofs}\label{paper-proofs}
We omit the proofs that have been checked by K. Memarian with
the $\coq$ proof assistant~\cite{Memarian10}.



\subsection{Notation}
Let $\act{t}$ be a family of reduction relations where 
$t$ ranges over the set of labels and $\epsilon$. Then we 
define:
\[
\wact{t} = \left\{
\begin{array}{ll}
(\act{\epsilon})^* &\mbox{if }t=\epsilon \\
(\act{\epsilon})^* \comp \act{t} \comp (\act{\epsilon})^*  &\mbox{otherwise}
\end{array}
\right.
\]
where as usual $R^*$ denote the reflexive and transitive closure of the relation $R$ 
and $\comp$ denotes the composition of relations.

\subsection{Proof of proposition \ref{soundness-small-step}} 
Given a $\vm$ code $C$, we define an `accessibility relation' 
$\access{C}$ as the least binary relation on $\set{0,\ldots,|C|-1}$ such that:

{\footnotesize
\[
\begin{array}{ll}

\infer{}
{i\access{C} i}
\qquad \qquad
&\infer{C[i]=\s{branch}(k)\quad (i+k+1)\access{C} j}
{i\access{C} j}

\end{array}
\]}

We also introduce a ternary relation $R(C,i,K)$ which relates 
a $\vm$ code $C$, a number $i\in\set{0,\ldots,|C|-1}$ and a continuation
$K$. 
The relation is 
defined as the least one that satisfies the following conditions.

{\footnotesize
\[
\begin{array}{ll}

\infer{
\begin{array}{c}
\\
i\access{C} j\qquad C[j]=\s{halt}
\end{array}}
{R(C,i,\s{halt})}

\qquad \qquad
&\infer{
\begin{array}{c}
i\access{C} i'\quad C=C_1 \cdot \cl{C}(S) \cdot C_2\\ 
i'=|C_1|\quad j=|C_1\cdot \cl{C}(S)| \quad R(C,j,K)
\end{array}}
{R(C,i,S\cdot K)}~.

\end{array}
\]}

The following properties are useful.

\begin{lemma}\label{access-lemma}
\Defitemf{(1)} The relation $\access{C}$ is transitive.

\Defitem{(2)}  If $i\access{C} j$ and $R(C,j,K)$ then $R(C,i,K)$.
\end{lemma}
The first property can be proven by induction on the definition of $\access{C}$ and  the second by induction on the structure of $K$. 

Next we can focus on the proposition.
The notation $C \stackrel{i}{\cdot} C'$ means that $i=|C|$.
Suppose that:

{\footnotesize
\[
\begin{array}{lll}
(S,K,s) \arrow (S',K',s') \quad (1)  &\text{and}  &R(C,i,S\cdot K) \quad (2)~.
\end{array}
\]}
From $(2)$, we know that there exist $i'$ and $i''$ such that:

{\footnotesize\[
\begin{array}{llll}
i \access{C} i' \quad (3),
&C=C_1\stackrel{i'}{\cdot}\cl{C}(S)\stackrel{i''}{\cdot}C_2 \quad (4), 
&\text{and} 
&R(C,i'',K) \quad (5)
\end{array}
\]}
and from $(3)$ it follows that:

{\footnotesize
\[ C\Gives (i,\sigma,s) \trarrow (i',\sigma,s) \quad (3')~.
\]}
We are looking for $j$ such that:

{\footnotesize
\[
\begin{array}{lll}
C \Gives (i,\sigma,s) \trarrow (j,\sigma,s') \quad (6), 
&\text{and} 
&R(C,j,S'\cdot K') \quad (7) ~.
\end{array}
\]}
We proceed by case analysis on $S$. We just detail the case of the conditional command as the
the remaining cases have similar proofs.
If $S=\s{if}$ $e_1<e_2$ $\s{then}$ $S_1$ $\s{else}$ $S_2$ then $(4)$ is rewritten as follows:

{\footnotesize  $$ C=C_1\stackrel{i'}{\cdot}\cl{C}(e_1)\cdot \cl{C}(e_2).\s{bge}(k_1)\stackrel{a}{\cdot}\cl{C}(S_1)\stackrel{b}{\cdot}\s{branch}(k_2) 
\stackrel{c}{\cdot} \cl{C}(S_2)\stackrel{i''}{\cdot}C_2~ $$}
where $c = a+k_1$ and $i''=c+k_2$.
  We distinguish two cases according to the evaluation of the boolean condition.
We describe the case 
$(e_1<e_2) \eval \s{true}$. We set $j=a$.
\begin{itemize}
	\item The instance of $(1)$ is $(S,K,s) \arrow (S_1,K,s)$.
	\item The reduction required in (6) takes the form 
$C \Gives (i,\sigma,s) \trarrow (i',\sigma,s) \trarrow (a,\sigma,s')$,
and it follows from $(3')$, the fact that $(e_1<e_2) \eval \s{true}$, and 
proposition \ref{soundness-big-step-prop}(2).

	\item Property $(7)$, follows from lemma \ref{access-lemma}(2), fact $(5)$, and 
the following proof tree:

{\footnotesize	  
$$ \infer{j\access{C} j \quad \infer{b \access{C} i'' \quad R(C,i'',K)}{R(C,b,K)}}{R(C,j,S_1\cdot K)} ~.$$ }
~\qed
\end{itemize}





\subsection{Proof of proposition \ref{well-formed-prop}}
We actually prove that for any expression $e$, statement $S$, and program $P$ the
following holds:

\Defitem{(1)} 
For any $n\in\Nat$ there is a unique $h$ such that 
$\cl{C}(e):h$, $h(0)=n$, and $h(|\cl{C}(e)|)=h(0)+1$.

\Defitem{(2)}
For any $S$, 
there is a unique $h$ such that $\cl{C}(S):h$, $h(0)=0$, and 
$h(|\cl{C}(e)|)=0$.

\Defitem{(3)}
There is a unique $h$ such that $\cl{C}(P):h$.

\subsection{Proof of proposition \ref{labelled-sim-imp-vm}} 

\Defitemf{(1)} By induction on the structure of the command $S$.

\Defitem{(2)} By iterating the following proposition.

\begin{proposition}
  If $(S,K,s) \stackrel{t}{\arrow} (S',K',s')$ and $R(C,i,S\cdot K)$ with $t=\ell$ or $t=\epsilon$ then 
  $C\Gives (i,\sigma,s) \wact{t} (j,\sigma,s')$ and
  $R(C,j,S'\cdot K')$.
\end{proposition}

This is an extension of proposition \ref{soundness-small-step} and it is proven in the same way with
an additional case for labelled commands. \qed

\subsection{Proof of proposition \ref{sim-vm-mips-prop}} 
\Proofitemf{(1)} The compilation of the $\vm$ instruction $\s{nop}(\ell)$ is the $\mips$ instruction $(\s{nop} \ \ell)$.

\Proofitem{(2)}				
By iterating the following proposition.

\begin{proposition}
Let $C : h$ be a well formed code.
If  $C \Gives (i,\sigma,s) \stackrel{t}\arrow (j,\sigma',s')$ with $t=\ell$ or $t=\epsilon$, $h(i) = |\sigma|$ and  $m \real \sigma,s$
then 
$\cl{C'}(C) \Gives (p(i,C),m) \wact{t} (p(j,C),m')$ and $m'\real \sigma',s'$.
\end{proposition} 

This is an extension of proposition \ref{simulation-vm-mips} and it is proven in the same way
with an additional case for the \s{nop} instruction.\qed

\subsection{Proof of proposition \ref{lab-instr-erasure-imp}} 
We extend the instrumentation to the continuations by defining:
\[
\cl{I}(S\cdot K) = \cl{I}(S)\cdot \cl{I}(K) \qquad
\cl{I}(\s{halt}) = \s{halt}~.
\]
Then we examine the possible reductions of a configuration
$(\cl{I}(S),\cl{I}(K),s[c/\w{cost}])$.

\begin{itemize}

\item
If $S$ is an unlabelled statement such as $\s{while} \ b \ \s{do} \ S'$
then $\cl{I}(S) = \s{while} \ b \ \s{do} \ \cl{I}(S')$ and 
assuming $(b,s) \eval \s{true}$ the reduction step is:
\[
(\cl{I}(S),\cl{I}(K),s[c/\w{cost}]) \arrow
(\cl{I}(S'), \cl{I}(S)\cdot \cl{I}(K),s[c/\w{cost}])~.
\]
Noticing that $\cl{I}(S)\cdot \cl{I}(K) = \cl{I}(S\cdot K)$,
this step is matched in the labelled language as follows:
\[
(S,K,s[c/\w{cost}]) \arrow
(S', S\cdot K,s[c/\w{cost}])~.
\]

\item On the other hand, if $S=\ell:S'$ is a labelled statement 
then $\cl{I}(S)= \w{inc}(\ell);\cl{I}(S')$ 
and, by a sequence of reductions steps, we have:
\[
(\cl{I}(S),\cl{I}(K),s[c/\w{cost}]) \trarrow
(\cl{I}(S'), \cl{I}(K),s[c+\kappa(\ell)/\w{cost}])~.
\]
This step is matched by the labelled reduction:
\[
(S,K,s[c/\w{cost}]) \act{\ell}
(S', K,s[c/\w{cost}])~.
\]
~\qed
\end{itemize}

\subsection{Proof of proposition \ref{global-commutation-prop}}
By diagram chasing using propositions \ref{labelled-sim-imp-vm}(1),
\ref{sim-vm-mips-prop}(1), and the definition \ref{labelling-def} of labelling. \qed

\subsection{Proof of proposition \ref{instrument-to-label-prop}}
Suppose that:
$$ (\cl{I}(\cl{L}(P)),s[c/\w{cost}]) \eval s'[c+\delta/\w{cost}] \text{ and } m\real s[c/\w{cost}]~. $$
Then, by proposition  \ref{lab-instr-erasure-imp}, for some $\lambda$:
$$ (\cl{L}(P),s[c/\w{cost}])\eval (s'[c/\w{cost}],\lambda) \text{ and } \kappa(\lambda)=\delta~. $$ 
Finally, by propositions  \ref{labelled-sim-imp-vm}(2) and \ref{sim-vm-mips-prop}(2) : 
$$ (\cl{C'}(\cl{C}(\cl{L}(P))),m) \eval (m',\lambda) \text{ and } m'\real s'[c/\w{cost}]~.$$
\qed

\subsection{Proof of proposition \ref{sound-label-prop}}
 If $\lambda=\ell_1 \cdots \ell_n$ then the 
computation is the concatenation of simple paths 
labelled with $\ell_1,\ldots,\ell_n$. Since $\kappa(\ell_i)$
bounds the cost of a simple path labelled with $\ell_i$, the
cost of the overall computation is bounded by $\kappa(\lambda)=
\kappa(\ell_1)+\cdots \kappa(\ell_n)$. \qed

\subsection{Proof of proposition \ref{precise-label-prop}}
Same proof as proposition \ref{sound-label-prop}, by replacing the word \emph{bounds} by \emph{is exactly} and the words \emph{bounded by} by \emph{exactly}. \qed

\subsection{Proof of proposition \ref{lab-sound}}  
In both labellings under consideration the root node is labelled.  An
obvious observation is that only commands of the shape \s{while} $b$
\s{do} $S$
introduce loops in the compiled code.  
We notice that both labelling introduce a label in the loop (though at different places).
Thus all loops go through a label and the compiled code is always sound.

To show the precision of the second labelling $\cl{L}_p$, we note the
following property. 

\begin{lemma}\label{precise-lemma}
A soundly labelled graph is precise if each label occurs at most once in the graph and 
if the immediate successors of the \s{bge} nodes are either \s{halt} (no successor) or labelled nodes.
\end{lemma}

Indeed, in a such a graph starting from a labelled node we can follow a unique path
up to a leaf, another labelled node, or a $\s{bge}$ node. In the last case, the 
hypotheses in the lemma \ref{precise-lemma} guarantee that the two 
simple paths one can follow from the $\s{bge}$ node have
the same length/cost.  \qed


\subsection{Proof of proposition \ref{ann-correct}} 
By applying consecutively proposition \ref{instrument-to-label-prop}
and propositions \ref{sound-label-prop} or \ref{precise-label-prop}. \qed

\newpage

\section{A $\C$ compiler}\label{C-compiler-sec}

This section gives an informal overview of the compiler, in particular it
highlights the main features of the intermediate languages, the purpose of the
compilation steps, and the optimisations. 

\subsection{$\Clight$} 
$\Clight$ is a large subset of the $\C$ language that we adopt as
the source language of our compiler. 
It features most of the types and operators
of $\C$. It includes pointer arithmetic, pointers to
functions, and \texttt{struct} and \texttt{union} types, as well as
all $\C$ control structures. The main difference with the $\C$
language is that $\Clight$ expressions are side-effect free, which
means that side-effect operators ($\codeex{=}$,$\codeex{+=}$,$\codeex{++}$,$\ldots$) and function calls
within expressions are not supported. 
Given a $\C$ program, we rely on the $\cil$ tool~\cite{CIL02} to deal 
with the idiosyncrasy of  $\C$ concrete syntax and to produce an
equivalent program in $\Clight$ abstract syntax.
We refer to  the $\compcert$ project~\cite{Leroy09} 
for a formal definition of the
$\Clight$ language. Here we just recall  in 
figure \ref{syntax-clight} its syntax which is 
classically structured in expressions, statements, functions, 
and whole programs. In order to limit the implementation effort,
our current compiler for $\Clight$ does {\em not} cover the operators
relating to the floating point type {\tt float}. So, in a nutshell,
the fragment of $\C$ we have implemented is $\Clight$ without
floating point.

\begin{figure}
  \label{syntax-clight}
  \footnotesize{
  \begin{tabular}{l l l l}
    Expressions: & $a$ ::= & $id$ 		& variable identifier \\
    & & $|$ $n$ 				& integer constant \\
    & & $|$ \texttt{sizeof}($\tau$) 		& size of a type \\
    & & $|$ $op_1$ $a$ 				& unary arithmetic operation \\
    & & $|$ $a$ $op_2$ $a$ 			& binary arithmetic operation \\
    & & $|$ $*a$ 				& pointer dereferencing \\
    & & $|$ $a.id$ 				& field access \\
    & & $|$ $\&a$ 				& taking the address of \\
    & & $|$ $(\tau)a$ 				& type cast \\
    & & $|$ $a ? a : a$ 			& conditional expression \\[10pt]
    Statements: & $s$ ::= & \texttt{skip} 	& empty statement \\
    & & $|$ $a=a$ 				& assignment \\
    & & $|$ $a=a(a^*)$ 			        & function call \\
    & & $|$ $a(a^*)$ 				& procedure call \\
    & & $|$ $s;s$ 				& sequence \\
    & & $|$ \texttt{if} $a$ \texttt{then} $s$ \texttt{else} $s$ & conditional \\
    & & $|$ \texttt{switch} $a$ $sw$ 		& multi-way branch \\
    & & $|$ \texttt{while} $a$ \texttt{do} $s$	& ``while'' loop \\
    & & $|$ \texttt{do} $s$ \texttt{while} $a$	& ``do'' loop \\
    & & $|$ \texttt{for}($s$,$a$,$s$) $s$	& ``for'' loop\\
    & & $|$ \texttt{break} 			& exit from current loop \\
    & & $|$ \texttt{continue} 			& next iteration of the current loop \\
    & & $|$ \texttt{return} $a^?$		& return from current function \\
    & & $|$ \texttt{goto} $lbl$			& branching \\
    & & $|$ $lbl$ : $s$				& labelled statement \\[10pt]
    Switch cases: & $sw$ ::= & \texttt{default} : $s$ 		& default case \\
    & & $|$ \texttt{case } $n$ : $s;sw$ 			& labelled case \\[10pt]
    Variable declarations: & $dcl$ ::= & $(\tau\quad id)^*$ 		& type and name\\[10pt]
    Functions: & $Fd$ ::= & $\tau$ $id(dcl)\{dcl;s\}$ 	& internal function \\
    & & $|$ \texttt{extern} $\tau$ $id(dcl)$ 			& external function\\[10pt]
    Programs: & $P$ ::= & $dcl;Fd^*;\texttt{main}=id$		& global variables, functions, entry point
  \end{tabular}}
  \caption{Syntax of the $\Clight$ language}
\end{figure}

\subsection{$\Cminor$}

$\Cminor$ is a simple, low-level imperative language, comparable to a
stripped-down, typeless variant of $\C$. Again we refer 
to the $\compcert$ project for its formal definition 
and we just recall in figure \ref{syntax-cminor}
its syntax which as for $\Clight$ is structured in
expressions, statements, functions, and whole programs.

\begin{figure}
  \label{syntax-cminor}
  \footnotesize{
  \begin{tabular}{l l l l}
    Signatures: & $sig$ ::= & \texttt{sig} $\vec{\texttt{int}}$ $(\texttt{int}|\texttt{void})$ & arguments and result \\[10pt]
    Expressions: & $a$ ::= & $id$ 		& local variable \\
     & & $|$ $n$ 				& integer constant \\
     & & $|$ \texttt{addrsymbol}($id$) 		& address of global symbol \\
     & & $|$ \texttt{addrstack}($\delta$) 	& address within stack data \\
     & & $|$ $op_1$ $a$				& unary arithmetic operation \\
     & & $|$ $op_2$ $a$ $a$			& binary arithmetic operation \\
     & & $|$ $\kappa[a]$			& memory read\\
     & & $|$ $a?a:a$			& conditional expression \\[10pt]
    Statements: & $s$ ::= & \texttt{skip} 	& empty statement \\
    & & $|$ $id=a$				& assignment \\
    & & $|$ $\kappa[a]=a$			& memory write \\
    & & $|$ $id^?=a(\vec{a}):sig$		& function call \\
    & & $|$ \texttt{tailcall} $a(\vec{a}):sig$	& function tail call \\
    & & $|$ \texttt{return}$(a^?)$		& function return \\
    & & $|$ $s;s$				& sequence \\
    & & $|$ \texttt{if} $a$ \texttt{then} $s$ \texttt{else} $s$		& conditional  \\
    & & $|$ \texttt{loop} $s$			& infinite loop \\
    & & $|$ \texttt{block} $s$		& block delimiting \texttt{exit} constructs \\
    & & $|$ \texttt{exit} $n$			& terminate the $(n+1)^{th}$ enclosing block \\
    & & $|$ \texttt{switch} $a$ $tbl$		& multi-way test and exit\\
    & & $|$ $lbl:s$				& labelled statement \\
    & & $|$ \texttt{goto} $lbl$			& jump to a label\\[10pt]
    Switch tables: & $tbl$ ::= & \texttt{default:exit}($n$) & \\
    & & $|$ \texttt{case} $i$: \texttt{exit}($n$);$tbl$ & \\[10pt]
    Functions: & $Fd$ ::= & \texttt{internal} $sig$ $\vec{id}$ $\vec{id}$ $n$ $s$ & internal function: signature, parameters, \\
    & & & local variables, stack size and body \\
    & & $|$ \texttt{external} $id$ $sig$ & external function \\[10pt]
    Programs: & $P$ ::= & \texttt{prog} $(id=data)^*$ $(id=Fd)^*$ $id$   & global variables, functions and entry point 
  \end{tabular}}
  \caption{Syntax of the $\Cminor$ language}
\end{figure}

\paragraph{Translation of $\Clight$ to $\Cminor$}
As in $\Cminor$ stack operations are made explicit, one has to know
which variables are stored in the stack. This 
information is produced by a static analysis that determines
the variables whose address may be `taken'. 
Also space is reserved for local arrays and structures. In a
second step, the proper compilation is performed: it consists mainly
in translating $\Clight$ control structures to the basic
ones available in $\Cminor$.

\subsection{$\RTLAbs$}

$\RTLAbs$ is the last architecture independent language in the
compilation process.  It is a rather straightforward {\em abstraction} of
the {\em architecture-dependent} $\RTL$ intermediate language
available in the $\compcert$ project and it is intended to factorize
some work common to the various target assembly languages
(e.g. optimizations) and thus to make retargeting of the compiler a
simpler matter.

We stress that in $\RTLAbs$ the structure of $\Cminor$ expressions
is lost and that this may have a negative impact on the following
instruction selection step. 
Still, the subtleties of instruction selection seem rather orthogonal
to our goals and we deem the possibility of retargeting
easily the compiler more important than the efficiency of the generated code.

\paragraph{Syntax.}
In $\RTLAbs$, programs are represented as \emph{control flow
  graphs} (CFGs for short). We associate with the nodes of the graphs 
instructions reflecting the $\Cminor$ commands. As usual, commands that change the control
flow of the program (e.g. loops, conditionals) are translated by inserting 
suitable branching instructions in the CFG.
The syntax of the language is depicted in table
\ref{RTLAbs:syntax}. Local variables are now represented by \emph{pseudo
  registers} that are available in unbounded number. The grammar rule $\w{op}$ that is
not detailed in table \ref{RTLAbs:syntax} defines usual arithmetic and boolean
operations (\codeex{+}, \codeex{xor}, \codeex{$\le$}, etc.) as well as constants
and conversions between sized integers.

\begin{table}
{\footnotesize
\[
\begin{array}{lllllll}
\w{return\_type} & ::= & \s{int} \Alt \s{void} & \qquad
\w{signature} & ::= & (\s{int} \rightarrow)^*\ \w{return\_type}\\
\end{array}
\]

\[
\begin{array}{lllllll}
\w{memq} & ::= & \s{int8s} \Alt \s{int8u} \Alt \s{int16s} \Alt \s{int16u}
                 \Alt \s{int32} & \qquad
\w{fun\_ref} & ::= & \w{fun\_name} \Alt \w{psd\_reg}
\end{array}
\]

\[
\begin{array}{llll}
\w{instruction} & ::= & \Alt \s{skip} \rightarrow \w{node} &
                        \quad \mbox{(no instruction)}\\
                &     & \Alt \w{psd\_reg} := \w{op}(\w{psd\_reg}^*)
                        \rightarrow \w{node} & \quad \mbox{(operation)}\\
                &     & \Alt \w{psd\_reg} := \s{\&}\w{var\_name}
                        \rightarrow \w{node} &
                        \quad \mbox{(address of a global)}\\
                &     & \Alt \w{psd\_reg} := \s{\&locals}[n]
                        \rightarrow \w{node} &
                        \quad \mbox{(address of a local)}\\
                &     & \Alt \w{psd\_reg} := \w{fun\_name}
                        \rightarrow \w{node} &
                        \quad \mbox{(address of a function)}\\
                &     & \Alt \w{psd\_reg} :=
                        \w{memq}(\w{psd\_reg}[\w{psd\_reg}])
                        \rightarrow \w{node} & \quad \mbox{(memory load)}\\
                &     & \Alt \w{memq}(\w{psd\_reg}[\w{psd\_reg}]) :=
                        \w{psd\_reg}
                        \rightarrow \w{node} & \quad \mbox{(memory store)}\\
                &     & \Alt \w{psd\_reg} := \w{fun\_ref}(\w{psd\_reg^*}) :
                        \w{signature} \rightarrow \w{node} &
                        \quad \mbox{(function call)}\\
                &     & \Alt \w{fun\_ref}(\w{psd\_reg^*}) : \w{signature}
                        & \quad \mbox{(function tail call)}\\
                &     & \Alt \s{test}\ \w{op}(\w{psd\_reg}^*) \rightarrow
                        \w{node}, \w{node} & \quad \mbox{(branch)}\\
                &     & \Alt \s{return}\ \w{psd\_reg}? & \quad \mbox{(return)}
\end{array}
\]

\[
\begin{array}{lll}
\w{fun\_def} & ::= & \w{fun\_name}(\w{psd\_reg}^*): \w{signature}\\
             &     & \s{result:} \w{psd\_reg}?\\
             &     & \s{locals:} \w{psd\_reg}^*\\
             &     & \s{stack:} n\\
             &     & \s{entry:} \w{node}\\
             &     & \s{exit:} \w{node}\\
             &     & (\w{node:} \w{instruction})^*
\end{array}
\]

\[
\begin{array}{lllllll}
\w{init\_datum} & ::= & \s{reserve}(n) \Alt \s{int8}(n) \Alt \s{int16}(n)
                       \Alt \s{int32}(n) & \qquad
\w{init\_data} & ::= & \w{init\_datum}^+
\end{array}
\]

\[
\begin{array}{lllllll}
\w{global\_decl} & ::= & \s{var}\ \w{var\_name} \w{\{init\_data\}} & \qquad
\w{fun\_decl} & ::= & \s{extern}\ \w{fun\_name} \w{(signature)} \Alt
                      \w{fun\_def}
\end{array}
\]

\[
\begin{array}{lll}
\w{program} & ::= & \w{global\_decl}^*\\
            &     & \w{fun\_decl}^*
\end{array}
\]}
\caption{Syntax of the $\RTLAbs$ language}\label{RTLAbs:syntax}
\end{table}

\paragraph{Translation of $\Cminor$ to $\RTLAbs$.}
Translating $\Cminor$ programs to $\RTLAbs$ programs mainly consists in
transforming $\Cminor$ commands in CFGs. Most commands are sequential and have a
rather straightforward linear translation. A conditional is translated in a
branch instruction; a loop is translated using a back edge in the CFG.

\subsection{$\RTL$}

As in $\RTLAbs$, the structure of $\RTL$ programs is based on CFGs. $\RTL$ is
the first architecture-dependant intermediate language of our compiler 
which, in its current version, targets the $\mips$ assembly language.

\paragraph{Syntax.}
$\RTL$ is very close to $\RTLAbs$. It is based on CFGs and explicits the $\mips$
instructions corresponding to the $\RTLAbs$ instructions. Type information
disappears: everything is represented using 32 bits integers. Moreover, each
global of the program is associated to an offset. The syntax of the language can
be found in table \ref{RTL:syntax}. The grammar rules $\w{unop}$, $\w{binop}$,
$\w{uncon}$, and $\w{bincon}$, respectively, represent the sets of unary
operations, binary operations, unary conditions and binary conditions of the
$\mips$ language.

\begin{table}
{\footnotesize
\[
\begin{array}{lllllll}
\w{size} & ::= & \s{Byte} \Alt \s{HalfWord} \Alt \s{Word}  & \qquad
\w{fun\_ref} & ::= & \w{fun\_name} \Alt \w{psd\_reg}
\end{array}
\]

\[
\begin{array}{llll}
\w{instruction} & ::= & \Alt \s{skip} \rightarrow \w{node} &
                        \quad \mbox{(no instruction)}\\
                &     & \Alt \w{psd\_reg} := n
                        \rightarrow \w{node} & \quad \mbox{(constant)}\\
                &     & \Alt \w{psd\_reg} := \w{unop}(\w{psd\_reg})
                        \rightarrow \w{node} & \quad \mbox{(unary operation)}\\
                &     & \Alt \w{psd\_reg} :=
                        \w{binop}(\w{psd\_reg},\w{psd\_reg})
                        \rightarrow \w{node} & \quad
                        \mbox{(binary operation)}\\
                &     & \Alt \w{psd\_reg} := \s{\&globals}[n]
                        \rightarrow \w{node} &
                        \quad \mbox{(address of a global)}\\
                &     & \Alt \w{psd\_reg} := \s{\&locals}[n]
                        \rightarrow \w{node} &
                        \quad \mbox{(address of a local)}\\
                &     & \Alt \w{psd\_reg} := \w{fun\_name}
                        \rightarrow \w{node} &
                        \quad \mbox{(address of a function)}\\
                &     & \Alt \w{psd\_reg} :=
                        \w{size}(\w{psd\_reg}[n])
                        \rightarrow \w{node} & \quad \mbox{(memory load)}\\
                &     & \Alt \w{size}(\w{psd\_reg}[n]) :=
                        \w{psd\_reg}
                        \rightarrow \w{node} & \quad \mbox{(memory store)}\\
                &     & \Alt \w{psd\_reg} := \w{fun\_ref}(\w{psd\_reg^*})
                        \rightarrow \w{node} &
                        \quad \mbox{(function call)}\\
                &     & \Alt \w{fun\_ref}(\w{psd\_reg^*})
                        & \quad \mbox{(function tail call)}\\
                &     & \Alt \s{test}\ \w{uncon}(\w{psd\_reg}) \rightarrow
                        \w{node}, \w{node} & \quad
                        \mbox{(branch unary condition)}\\
                &     & \Alt \s{test}\ \w{bincon}(\w{psd\_reg},\w{psd\_reg})
                        \rightarrow \w{node}, \w{node} & \quad
                        \mbox{(branch binary condition)}\\
                &     & \Alt \s{return}\ \w{psd\_reg}? & \quad
                        \mbox{(return)}
\end{array}
\]

\[
\begin{array}{lllllll}
\w{fun\_def} & ::= & \w{fun\_name}(\w{psd\_reg}^*) & \qquad
\w{program} & ::= & \s{globals}: n\\
             &     & \s{result:} \w{psd\_reg}? & \qquad
            &     & \w{fun\_def}^*\\
             &     & \s{locals:} \w{psd\_reg}^*\\
             &     & \s{stack:} n\\
             &     & \s{entry:} \w{node}\\
             &     & \s{exit:} \w{node}\\
             &     & (\w{node:} \w{instruction})^*
\end{array}
\]}
\caption{Syntax of the $\RTL$ language}\label{RTL:syntax}
\end{table}

\paragraph{Translation of $\RTLAbs$ to $\RTL$.}
This translation is mostly straightforward. A $\RTLAbs$ instruction is often
directly translated to a corresponding $\mips$ instruction. There are a few
exceptions: some $\RTLAbs$ instructions are expanded in two or more $\mips$
instructions. When the translation of a $\RTLAbs$ instruction requires more than
a few simple $\mips$ instruction, it is translated into a call to a function
defined in the preamble of the compilation result.

\subsection{$\ERTL$}

As in $\RTL$, the structure of $\ERTL$ programs is based on CFGs. $\ERTL$
explicits the calling conventions of the $\mips$ assembly language.

\paragraph{Syntax.}
The syntax of the language is given in table \ref{ERTL:syntax}. The main
difference between $\RTL$ and $\ERTL$ is the use of hardware registers.
Parameters are passed in specific hardware registers; if there are too many
parameters, the remaining are stored in the stack. Other conventionally specific
hardware registers are used: a register that holds the result of a function, a
register that holds the base address of the globals, a register that holds the
address of the top of the stack, and some registers that need to be saved when
entering a function and whose values are restored when leaving a
function. Following these conventions, function calls do not list their
parameters anymore; they only mention their number. Two new instructions appear
to allocate and deallocate on the stack some space needed by a function to
execute. Along with these two instructions come two instructions to fetch or
assign a value in the parameter sections of the stack; these instructions cannot
yet be translated using regular load and store instructions because we do not
know the final size of the stack area of each function. At last, the return
instruction has a boolean argument that tells whether the result of the function
may later be used or not (this is exploited for optimizations).

\begin{table}
{\footnotesize
\[
\begin{array}{lllllll}
\w{size} & ::= & \s{Byte} \Alt \s{HalfWord} \Alt \s{Word}  & \qquad
\w{fun\_ref} & ::= & \w{fun\_name} \Alt \w{psd\_reg}
\end{array}
\]

\[
\begin{array}{llll}
\w{instruction} & ::= & \Alt \s{skip} \rightarrow \w{node} &
                        \quad \mbox{(no instruction)}\\
                &     & \Alt \s{NewFrame} \rightarrow \w{node} &
                        \quad \mbox{(frame creation)}\\
                &     & \Alt \s{DelFrame} \rightarrow \w{node} &
                        \quad \mbox{(frame deletion)}\\
                &     & \Alt \w{psd\_reg} := \s{stack}[\w{slot}, n]
                        \rightarrow \w{node} &
                        \quad \mbox{(stack load)}\\
                &     & \Alt \s{stack}[\w{slot}, n] := \w{psd\_reg}
                        \rightarrow \w{node} &
                        \quad \mbox{(stack store)}\\
                &     & \Alt \w{hdw\_reg} := \w{psd\_reg}
                        \rightarrow \w{node} &
                        \quad \mbox{(pseudo to hardware)}\\
                &     & \Alt \w{psd\_reg} := \w{hdw\_reg}
                        \rightarrow \w{node} &
                        \quad \mbox{(hardware to pseudo)}\\
                &     & \Alt \w{psd\_reg} := n
                        \rightarrow \w{node} & \quad \mbox{(constant)}\\
                &     & \Alt \w{psd\_reg} := \w{unop}(\w{psd\_reg})
                        \rightarrow \w{node} & \quad \mbox{(unary operation)}\\
                &     & \Alt \w{psd\_reg} :=
                        \w{binop}(\w{psd\_reg},\w{psd\_reg})
                        \rightarrow \w{node} & \quad
                        \mbox{(binary operation)}\\
                &     & \Alt \w{psd\_reg} := \w{fun\_name}
                        \rightarrow \w{node} &
                        \quad \mbox{(address of a function)}\\
                &     & \Alt \w{psd\_reg} :=
                        \w{size}(\w{psd\_reg}[n])
                        \rightarrow \w{node} & \quad \mbox{(memory load)}\\
                &     & \Alt \w{size}(\w{psd\_reg}[n]) :=
                        \w{psd\_reg}
                        \rightarrow \w{node} & \quad \mbox{(memory store)}\\
                &     & \Alt \w{fun\_ref}(n) \rightarrow \w{node} &
                        \quad \mbox{(function call)}\\
                &     & \Alt \w{fun\_ref}(n)
                        & \quad \mbox{(function tail call)}\\
                &     & \Alt \s{test}\ \w{uncon}(\w{psd\_reg}) \rightarrow
                        \w{node}, \w{node} & \quad
                        \mbox{(branch unary condition)}\\
                &     & \Alt \s{test}\ \w{bincon}(\w{psd\_reg},\w{psd\_reg})
                        \rightarrow \w{node}, \w{node} & \quad
                        \mbox{(branch binary condition)}\\
                &     & \Alt \s{return}\ b & \quad \mbox{(return)}
\end{array}
\]

\[
\begin{array}{lllllll}
\w{fun\_def} & ::= & \w{fun\_name}(n) & \qquad
\w{program} & ::= & \s{globals}: n\\
             &     & \s{locals:} \w{psd\_reg}^* & \qquad
            &     & \w{fun\_def}^*\\
             &     & \s{stack:} n\\
             &     & \s{entry:} \w{node}\\
             &     & (\w{node:} \w{instruction})^*
\end{array}
\]}
\caption{Syntax of the $\ERTL$ language}\label{ERTL:syntax}
\end{table}

\paragraph{Translation of $\RTL$ to $\ERTL$.}
The work consists in expliciting the conventions previously mentioned. These
conventions appear when entering, calling and leaving a function, and when
referencing a global variable or the address of a local variable.

\paragraph{Optimizations.}
A \emph{liveness analysis} is performed on $\ERTL$ to replace unused
instructions by a $\s{skip}$. An instruction is tagged as unused when it
performs an assignment on a register that will not be read afterwards. Also, the
result of the liveness analysis is exploited by a \emph{register allocation}
algorithm whose result is to efficiently associate a physical location (a
hardware register or an address in the stack) to each pseudo register of the
program.

\subsection{$\LTL$}

As in $\ERTL$, the structure of $\LTL$ programs is based on CFGs. Pseudo
registers are not used anymore; instead, they are replaced by physical locations
(a hardware register or an address in the stack).

\paragraph{Syntax.}
Except for a few exceptions, the instructions of the language are those of
$\ERTL$ with hardware registers replacing pseudo registers. Calling and
returning conventions were explicited in $\ERTL$; thus, function calls and
returns do not need parameters in $\LTL$. The syntax is defined in table
\ref{LTL:syntax}.

\begin{table}
{\footnotesize
\[
\begin{array}{lllllll}
\w{size} & ::= & \s{Byte} \Alt \s{HalfWord} \Alt \s{Word}  & \qquad
\w{fun\_ref} & ::= & \w{fun\_name} \Alt \w{hdw\_reg}
\end{array}
\]

\[
\begin{array}{llll}
\w{instruction} & ::= & \Alt \s{skip} \rightarrow \w{node} &
                        \quad \mbox{(no instruction)}\\
                &     & \Alt \s{NewFrame} \rightarrow \w{node} &
                        \quad \mbox{(frame creation)}\\
                &     & \Alt \s{DelFrame} \rightarrow \w{node} &
                        \quad \mbox{(frame deletion)}\\
                &     & \Alt \w{hdw\_reg} := n
                        \rightarrow \w{node} & \quad \mbox{(constant)}\\
                &     & \Alt \w{hdw\_reg} := \w{unop}(\w{hdw\_reg})
                        \rightarrow \w{node} & \quad \mbox{(unary operation)}\\
                &     & \Alt \w{hdw\_reg} :=
                        \w{binop}(\w{hdw\_reg},\w{hdw\_reg})
                        \rightarrow \w{node} & \quad
                        \mbox{(binary operation)}\\
                &     & \Alt \w{hdw\_reg} := \w{fun\_name}
                        \rightarrow \w{node} &
                        \quad \mbox{(address of a function)}\\
                &     & \Alt \w{hdw\_reg} := \w{size}(\w{hdw\_reg}[n])
                        \rightarrow \w{node} & \quad \mbox{(memory load)}\\
                &     & \Alt \w{size}(\w{hdw\_reg}[n]) := \w{hdw\_reg}
                        \rightarrow \w{node} & \quad \mbox{(memory store)}\\
                &     & \Alt \w{fun\_ref}() \rightarrow \w{node} &
                        \quad \mbox{(function call)}\\
                &     & \Alt \w{fun\_ref}()
                        & \quad \mbox{(function tail call)}\\
                &     & \Alt \s{test}\ \w{uncon}(\w{hdw\_reg}) \rightarrow
                        \w{node}, \w{node} & \quad
                        \mbox{(branch unary condition)}\\
                &     & \Alt \s{test}\ \w{bincon}(\w{hdw\_reg},\w{hdw\_reg})
                        \rightarrow \w{node}, \w{node} & \quad
                        \mbox{(branch binary condition)}\\
                &     & \Alt \s{return} & \quad \mbox{(return)}
\end{array}
\]

\[
\begin{array}{lllllll}
\w{fun\_def} & ::= & \w{fun\_name}(n) & \qquad
\w{program} & ::= & \s{globals}: n\\
             &     & \s{locals:} n & \qquad
            &     & \w{fun\_def}^*\\
             &     & \s{stack:} n\\
             &     & \s{entry:} \w{node}\\
             &     & (\w{node:} \w{instruction})^*
\end{array}
\]}
\caption{Syntax of the $\LTL$ language}\label{LTL:syntax}
\end{table}

\paragraph{Translation of $\ERTL$ to $\LTL$.} The translation relies on the
results of the liveness analysis and of the register allocation. Unused
instructions are eliminated and each pseudo register is replaced by a physical
location. In $\LTL$, the size of the stack frame of a function is known;
instructions intended to load or store values in the stack are translated
using regular load and store instructions.

\paragraph{Optimizations.} A \emph{graph compression} algorithm removes empty
instructions generated by previous compilation passes and by the liveness
analysis.

\subsection{$\LIN$}

In $\LIN$, the structure of a program is no longer based on CFGs. Every function
is represented as a sequence of instructions.

\paragraph{Syntax.}
The instructions of $\LIN$ are very close to those of $\LTL$. \emph{Program
  labels}, \emph{gotos} and branch instructions handle the changes in the
control flow. The syntax of $\LIN$ programs is shown in table \ref{LIN:syntax}.

\begin{table}
{\footnotesize
\[
\begin{array}{lllllll}
\w{size} & ::= & \s{Byte} \Alt \s{HalfWord} \Alt \s{Word}  & \qquad
\w{fun\_ref} & ::= & \w{fun\_name} \Alt \w{hdw\_reg}
\end{array}
\]

\[
\begin{array}{llll}
\w{instruction} & ::= & \Alt \s{NewFrame} &
                        \quad \mbox{(frame creation)}\\
                &     & \Alt \s{DelFrame} &
                        \quad \mbox{(frame deletion)}\\
                &     & \Alt \w{hdw\_reg} := n & \quad \mbox{(constant)}\\
                &     & \Alt \w{hdw\_reg} := \w{unop}(\w{hdw\_reg})
                        & \quad \mbox{(unary operation)}\\
                &     & \Alt \w{hdw\_reg} :=
                        \w{binop}(\w{hdw\_reg},\w{hdw\_reg})
                        & \quad \mbox{(binary operation)}\\
                &     & \Alt \w{hdw\_reg} := \w{fun\_name}
                        & \quad \mbox{(address of a function)}\\
                &     & \Alt \w{hdw\_reg} := \w{size}(\w{hdw\_reg}[n])
                        & \quad \mbox{(memory load)}\\
                &     & \Alt \w{size}(\w{hdw\_reg}[n]) := \w{hdw\_reg}
                        & \quad \mbox{(memory store)}\\
                &     & \Alt \s{call}\ \w{fun\_ref}
                        & \quad \mbox{(function call)}\\
                &     & \Alt \s{tailcall}\ \w{fun\_ref}
                        & \quad \mbox{(function tail call)}\\
                &     & \Alt \w{uncon}(\w{hdw\_reg}) \rightarrow
                        \w{node} & \quad
                        \mbox{(branch unary condition)}\\
                &     & \Alt \w{bincon}(\w{hdw\_reg},\w{hdw\_reg})
                        \rightarrow \w{node} & \quad
                        \mbox{(branch binary condition)}\\
                &     & \Alt \w{mips\_label:} & \quad \mbox{($\mips$ label)}\\
                &     & \Alt \s{goto}\ \w{mips\_label} & \quad \mbox{(goto)}\\
                &     & \Alt \s{return} & \quad \mbox{(return)}
\end{array}
\]

\[
\begin{array}{lllllll}
\w{fun\_def} & ::= & \w{fun\_name}(n) & \qquad
\w{program} & ::= & \s{globals}: n\\
             &     & \s{locals:} n & \qquad
            &     & \w{fun\_def}^*\\
             &     & \w{instruction}^*
\end{array}
\]}
\caption{Syntax of the $\LIN$ language}\label{LIN:syntax}
\end{table}

\paragraph{Translation of $\LTL$ to $\LIN$.}
This translation amounts to transform in an efficient way the graph structure of
functions into a linear structure of sequential instructions.

\subsection{$\mips$}

$\mips$ is a rather simple assembly language. As for other assembly languages, a
program in $\mips$ is a sequence of instructions. The $\mips$ code produced by
the compilation of a $\Clight$ program starts with a preamble in which some
useful and non-primitive functions are predefined (e.g. conversion from 8 bits
unsigned integers to 32 bits integers). The subset of the $\mips$ assembly
language that the compilation produces is defined in table \ref{mips:syntax}.

\begin{table}
{\footnotesize
\[
\begin{array}{lllllllllll}
\w{load} & ::= & \s{lb} \Alt \s{lhw} \Alt \s{lw} & \qquad
\w{store} & ::= & \s{sb} \Alt \s{shw} \Alt \s{sw} & \qquad
\w{fun\_ref} & ::= & \w{fun\_name} \Alt \w{hdw\_reg}
\end{array}
\]

\[
\begin{array}{llll}
\w{instruction} & ::= & \Alt \s{nop} & \quad \mbox{(empty instruction)}\\
                &     & \Alt \s{li}\ \w{hdw\_reg}, n & \quad \mbox{(constant)}\\
                &     & \Alt \w{unop}\ \w{hdw\_reg}, \w{hdw\_reg}
                        & \quad \mbox{(unary operation)}\\
                &     & \Alt \w{binop}\
                        \w{hdw\_reg},\w{hdw\_reg},\w{hdw\_reg}
                        & \quad \mbox{(binary operation)}\\
                &     & \Alt \s{la}\ \w{hdw\_reg}, \w{fun\_name}
                        & \quad \mbox{(address of a function)}\\
                &     & \Alt \w{load}\ \w{hdw\_reg}, n(\w{hdw\_reg})
                        & \quad \mbox{(memory load)}\\
                &     & \Alt \w{store}\ \w{hdw\_reg}, n(\w{hdw\_reg})
                        & \quad \mbox{(memory store)}\\
                &     & \Alt \s{call}\ \w{fun\_ref}
                        & \quad \mbox{(function call)}\\
                &     & \Alt \w{uncon}\ \w{hdw\_reg}, \w{node} & \quad
                        \mbox{(branch unary condition)}\\
                &     & \Alt \w{bincon}\ \w{hdw\_reg},\w{hdw\_reg},\w{node}
                        & \quad \mbox{(branch binary condition)}\\
                &     & \Alt \w{mips\_label:} & \quad \mbox{($\mips$ label)}\\
                &     & \Alt \s{j}\ \w{mips\_label} & \quad \mbox{(goto)}\\
                &     & \Alt \s{return} & \quad \mbox{(return)}
\end{array}
\]

\[
\begin{array}{lll}
\w{program} & ::= & \s{globals}: n\\
            &     & \s{entry}: \w{mips\_label}^*\\
            &     & \w{instruction}^*
\end{array}
\]}
\caption{Syntax of the $\mips$ language}\label{mips:syntax}
\end{table}

\paragraph{Translation of $\LIN$ to $\mips$.} This final translation is simple
enough. Stack allocation and deallocation are explicited and the function
definitions are sequentialized.

\subsection{Benchmarks}
\label{C-compiler-benchmarks}
\begin{figure}

\begin{center}
\begin{tabular}{r|rrr}
& \texttt{gcc -O0} & \texttt{acc} & \texttt{gcc -O1} \\
\hline
badsort & 55.93 & 34.51 & 12.96 \\
fib & 76.24 & 34.28 & 45.68 \\
mat\_det & 163.42 & 156.20 & 54.76 \\ 
min & 12.21 & 16.25 & 3.95 \\
quicksort & 27.46 & 17.95 & 9.41 \\
search & 463.19 & 623.79 & 155.38 \\
\hline
\end{tabular}
\end{center}

\caption{Benchmarks results (execution time is given in seconds).}
\label{fig:benchmark-results}
\end{figure}

To ensure that our prototype compiler is realistic, we performed some
preliminary benchmarks on a 183MHz MIPS 4KEc processor, running a
linux based distribution. We compared the wall clock execution time of
several simple~\C\ programs compiled with our compiler against the ones
produced by {\sc Gcc} set up with optimization levels 0 and 1. As
shown by Figure~\ref{fig:benchmark-results}, our prototype compiler
produces executable programs that are on average faster than 
{\sc Gcc}'s without optimizations.


{\footnotesize
\begin{thebibliography}{99}

\bibitem{AbsInt}
AbsInt Angewandte Informatik. {\tt http://www.absint.com/}.

\bibitem{CercoDeliverable}
R.M.~Amadio, N.~Ayache, K.~Memarian, R.~Saillard, Y.~R\'egis-Gianas.
\newblock Compiler Design and Intermediate Languages.
\newblock Deliverable 2.1 of~\cite{Cerco10}.

\bibitem{Cerco10}
Certified Complexity (Project description).
\newblock  ICT-2007.8.0 FET Open, Grant 243881. 
{\tt http://cerco.cs.unibo.it}.

\bibitem{SCADE}
Esterel Technologies. 
{\tt http://www.esterel-technologies.com}.

\bibitem{Frama-C}
$\framac$ software analysers.
{\tt http://frama-c.com/}.

\bibitem{AbsintScade}
C.~Ferdinand, R.~Heckmann, T.~ Le Sergent, D.~Lopes, B.~Martin, X.~Fornari, and F.~Martin.
\newblock Combining a high-level design tool for safety-critical
systems with a tool for {WCET} analysis of executables.
\newblock  In 
{\em Embedded Real Time Software (ERTS)}, 
2008.

\bibitem{Fornari10}
X.~Fornari.
\newblock Understanding how {SCADE} suite {KCG} generates safe {C} code.
\newblock White paper, Esterel Technologies, 2010.


\bibitem{Larus05}
J.~Larus.
\newblock Assemblers, linkers, and the SPIM simulator.
\newblock Appendix of {\em Computer Organization and Design: the hw/sw interface}, 
by Hennessy and Patterson, 2005.


\bibitem{Leroy09}
X.~Leroy.
\newblock Formal verification of a realistic compiler.
\newblock {\em Commun. ACM}, 52(7):107-115, 2009.


\bibitem{Leroy09-ln}
X.~Leroy.
\newblock Mechanized semantics, with applications to program proof and compiler verification.
\newblock {\em Marktoberdorf summer school}, 2009. 


\bibitem{MP67}
J.~McCarthy and J.~Painter.
\newblock Correctness of a compiler for arithmetic expressions.
\newblock In {\em Math. aspects of Comp. Sci. 1}, vol. 19
of Symp. in Appl. Math., AMS, 1967.

\bibitem{Memarian10}
K.~Memarian.
\newblock Complexit\'e Certifi\'ee.
\newblock {\em Travail d'\'etude et de recherche}, 
Master Informatique, Universit\'e Paris Diderot, 2010.\\
\url{http://www.pps.jussieu.fr/~yrg/miniCerCo/}

\bibitem{CIL02}
 G.~Necula, S.~McPeak, S.P.~Rahul, and W.~Weimer.
\newblock CIL: Intermediate Language and Tools for Analysis and Transformation of C Programs.
\newblock In {\em Proceedings of Conference on Compiler Construction}, 
Springer LNCS 2304:213--228, 2002.

\bibitem{Pottier}
F.~Pottier.
\newblock Compilation (INF 564), \'Ecole Polytechnique, 2009-2010. 
{\tt http://www.enseignement.polytechnique.fr/informatique/INF564/}. 

\bibitem{W09}
R.~Wilhelm et al.
\newblock The worst-case execution-time problem - overview of methods and survey of tools. 
\newblock {\em ACM Trans. Embedded Comput. Syst.}, 7(3), 2008.

\end{thebibliography}}
\end{document}